\documentstyle[12pt]{article}   

\large
\newcommand{\beq}{\begin{equation}}
\newcommand{\eeq}{\end{equation}}

\makeatletter
\@addtoreset{equation}{section}
\makeatother

\newtheorem{Theorem}{Theorem}[section]
\newtheorem{Definition}{Definition}[section]
\newtheorem{Lemma}{Lemma}[section]

\def\be{\begin{equation}}
\def\ee{\end{equation}}
\def\ba{\begin{eqnarray}}
\def\ea{\end{eqnarray}}

\def\a{{\cal A}}

\def\ab{{\overline{{\cal A}}}}

\def\Co{{\mathchoice
{\setbox0=\hbox{$\displaystyle\rm C$}\hbox{\hbox to0pt
{\kern0.4\wd0\vrule height0.9\ht0\hss}\box0}}
{\setbox0=\hbox{$\textstyle\rm C$}\hbox{\hbox to0pt
{\kern0.4\wd0\vrule height0.9\ht0\hss}\box0}}
{\setbox0=\hbox{$\scriptstyle\rm C$}\hbox{\hbox to0pt
{\kern0.4\wd0\vrule height0.9\ht0\hss}\box0}}
{\setbox0=\hbox{$\scriptscriptstyle\rm C$}\hbox{\hbox to0pt
{\kern0.4\wd0\vrule height0.9\ht0\hss}\box0}}}}
\def\Rl{{\mathchoice
{\setbox0=\hbox{$\displaystyle\rm R$}\hbox{\hbox to0pt
{\kern0.4\wd0\vrule height0.9\ht0\hss}\box0}}
{\setbox0=\hbox{$\textstyle\rm R$}\hbox{\hbox to0pt
{\kern0.4\wd0\vrule height0.9\ht0\hss}\box0}}
{\setbox0=\hbox{$\scriptstyle\rm R$}\hbox{\hbox to0pt
{\kern0.4\wd0\vrule height0.9\ht0\hss}\box0}}
{\setbox0=\hbox{$\scriptscriptstyle\rm R$}\hbox{\hbox to0pt
{\kern0.4\wd0\vrule height0.9\ht0\hss}\box0}}}}

\title{Quantum Spin Dynamics (QSD) : VII.\\ Symplectic Structures and 
Continuum Lattice Formulations of Gauge Field Theories}
\author{T. Thiemann\thanks{thiemann@aei-potsdam.mpg.de} \\
       MPI f. Gravitationsphysik, Albert-Einstein-Institut,\\ 
           Am M\"uhlenberg 1, 14476 Golm near Potsdam, Germany} 
\date{{\small Preprint AEI-2000-026}} 

\begin{document}

\maketitle

\begin{abstract}
Interesting non-linear functions on the phase spaces of classical field 
theories can never be quantized immediately because the basic
fields of the theory become operator valued distributions. Therefore,
one is usually forced to find a classical substitute for such a function
depending on a regulator which is expressed in terms of smeared quantities 
and which can be quantized in a well-defined way. Namely, the smeared 
functions define a new symplectic manifold of their own which is easy to 
quantize. Finally one must remove the regulator and establish that the 
final operator, if it exists, has the correct classical limit.

In this paper we investigate these steps for diffeomorphism invariant 
quantum field theories of connections. We introduce a generalized 
projective family of 
symplectic manifolds, coordinatized by the smeared fields, which is
labelled by a pair consisting of a graph and 
another graph dual to it. We show that there exists a generalized  
projective sequence 
of symplectic manifolds whose limit agrees with the symplectic manifold that 
one started from. 

This family of symplectic manifolds is easy to quantize and we illustrate 
the programme outlined above by applying it to the Gauss constraint.
The framework developed here is the classical cornerstone on which
the semi-classical analysis developed in a new series of papers called  
``Gauge Theory Coherent States'' is based. 

This article also complements, as a side result,
earlier work by Ashtekar, Corichi and Zapata who observed that 
certain operators are non-commuting on certain states although the 
Poisson brackets between the classical functions that these authors 
based the quantization on, vanish. We show that there are other functions 
on the classical phase space which give rise to the same operators but
whose Poisson algebra precisely mirrors the quantum commutator algebra.
\end{abstract}

\section{Introduction}
\label{s1}

Quantum General Relativity (QGR) has matured over the past decade to a 
mathematically well-defined theory of quantum gravity. 
In contrast to string theory, by definition, GQR is a
manifestly background independent, diffeomorphism 
invariant and non-perturbative theory.
The obvious advantage is that one will never have to postulate the
existence of a non-perturbative extension of the theory,
which in string theory has been called the still unknown 
M(ystery)-Theory.

The disadvantage of a non-perturbative and background independent
formulation is, of course, that one is faced with new and interesting 
mathematical problems so that one cannot just go ahead and 
``start calculating scattering amplitudes'': 
As there is no background around which one could perturb, rather the full 
metric is fluctuating, one is not
doing quantum field theory on a spacetime but only on a differential
manifold. Once there is no (Minkowski) metric at our disposal, one loses
familiar notions such as causality structure, locality, Poincar\'e group 
and so forth, in other words, the theory is not a theory to which
the Wightman axioms apply. Therefore, one must build an entirely
new mathematical apparatus to treat the resulting quantum field theory 
which is {\it drastically different from the Fock space picture 
to which particle physicists are used to}.

As a consequence, the mathematical formulation of the theory was the main 
focus of research in the field over the past decade. The main 
achievements to date are the following (more or less in chronological 
order) : 
\begin{itemize}
\item[i)] {\it Kinematical Framework}\\
The starting 
point was the introduction of new field variables \cite{1} for the 
gravitational field which are better suited to a background  
independent formulation of the quantum theory than the ones employed
until that time. In its original version these variables were
complex valued, however, currently their real valued version,  
considered first in \cite{1a} for {\it classical} Euclidean gravity and 
later in \cite{1b} for {\it classical} Lorentzian gravity, is preferred 
because 
to date it seems that it is only with these variables that one can rigorously
define the dynamics of Euclidean or Lorentzian  {\it quantum} gravity
\cite{1c}. \\
These variables are coordinates for the infinite dimensional phase
space of an $SU(2)$ gauge theory subject to further constraints 
besides the Gauss law, that is, a connection and a canonically
conjugate electric field. As such, it is very natural to introduce
smeared functions of these variables, specifically Wilson loop and 
electric flux functions. (Notice that one does not need a metric 
to define these functions, that is, they are background independent).
This had been done for ordinary gauge fields already before in \cite{2} 
and was then reconsidered for gravity (see e.g. \cite{3}).\\
The next step was the choice of a representation of the canonical
commutation relations between the electric and magnetic degrees
of freedom. This involves the choice of a suitable space of 
distributional connections \cite{4} and a faithful measure thereon \cite{5}
which, as one can show \cite{6}, is $\sigma$-additive.
The proof that the resulting Hilbert space indeed solves the adjointness 
relations induced by the reality structure of the classical theory
as well as the canonical commutation relations induced by the symplectic 
structure of the classical theory can be found in \cite{7}.
Independently, a second representation, called the loop 
representation, of the canonical commutation
relations had been advocated (see e.g. \cite{8} and especially \cite{8a}
and references therein)
but both representations were shown to be unitarily equivalent in
\cite{9} (see also \cite{10} for a different method of proof).\\
This is then the first major achievement : The theory is based on
a rigorously defined kinematical framework.
\item[ii)] {\it Geometrical Operators}\\
The second major achievement concerns the spectra of positive 
semi-definite, self-adjoint geometrical
operators measuring lengths \cite{11}, areas \cite{12,13}
and volumes \cite{12,14,15,16,8} of curves, surfaces and regions
in spacetime. These spectra are pure point (discete) and imply a discrete
Planck scale structure. It should be pointed out that the discreteness
is, in contrast to other approaches to quantum gravity, not put in
by hand but it is a {\it prediction} !
\item[iii)] {\it Regularization- and Renormalization Techniques}\\
The third major achievement is that there is a new 
regularization and renormalization technique \cite{17,18}
for diffeomorphism covariant, density-one-valued operators at our disposal
which was successfully tested in model theories \cite{19}. This
technique can be applied, in particular, to the standard model
coupled to gravity \cite{20,21} and to the Poincar\'e generators at 
spatial infinity \cite{22}. In particular, it works for {\it Lorentzian}
gravity while all earlier proposals could at best work in the Euclidean 
context only (see, e.g. \cite{8a} and references therein).
The algebra of important operators of the
resulting quantum field theories was shown to be consistent \cite{23}. 
Most surprisingly, these operators are {\it UV and IR finite} !
Notice that this result, at least as far as these operators are 
concerned, is stronger 
than the believed but unproved finiteness of scattering amplitudes
order by order in perturbation theory of the five critical
string theories, figuratively speaking, we claim that our perturbation series 
converges.
The absence of the divergences that usually plague interacting quantum fields
propagating on a Minkowski background can be understood intuitively
from the diffeomorphism invariance of the theory : ``short and long distances
are gauge equivalent''. We will elaborate more on this point in future 
publications. The classical limit of the above mentioned operators will
be studied in our companion paper \cite{23f}.
\item[iv)] {\it Spin Foam Models}\\
After the construction of the densely defined Hamiltonian constraint
operator of \cite{17,18}, a formal, Euclidean functional integral was
constructed in \cite{23a} and gave rise to the so-called spin foam 
models   
(a spin foam is a history of a graph with faces as the history of 
edges)
\cite{23b}. Spin foam models are in close connection with causal
spin-network evolutions \cite{23c}, state sum models \cite{23d} and
topological quantum field theory, in particular BF theory 
\cite{23e}. To
date most results are at a formal level and for the Euclidean 
version of the
theory only but the programme is exciting since it may restore 
manifest
four-dimensional diffeomorphism invariance which in the Hamiltonian
formulation is somewhat hidden.
\item[v)]
Finally, the fifth major achievement is the existence of a rigorous and 
satisfactory framework \cite{24,25,26,27,28,29,30} for the quantum 
statistical description of black holes
which reproduces the Bekenstein-Hawking Entropy-Area relation and applies,
in particular, to physical Schwarzschild black holes while stringy black 
holes so far are under control only for extremal charged black holes.
\end{itemize}
Summarizing, the work of the past decade has now 
culminated in a promising starting point for a quantum theory of the 
gravitational field plus matter and the stage is set to address physical 
questions. In particular, one would like to make contact with the language
that particle physicists are more familiar with, that is, perturbation 
theory. In other words, one should be able to define something like
{\it gravitons and photons propagating on a fluctuating quantum spacetime}.
By this we mean the following :\\
Suppose we want to study the semi-classical limit of our quantum gravity 
theory, that is, a limit in which the gravitational field behaves 
almost classical. This does not mean that we want to treat gravity as 
a background field \cite{31}, rather we take all the quantum fluctuations
into account but try to find a state with respect to which those fluctuations
(around the Minkowski metric) are minimal. With respect to such a 
``background state'' one can study relative excitations of the 
gravitational field (gravitons) or of matter fields (such as photons).

In order to do this we must first develop an appropriate semi-classical
framework which we will do in \cite{46,47,45}. But even before doing this 
we must examine the following issue which seems not to have been 
sufficiently appreciated throughout the literature so far :\\
Namely, the quantum theory is based on certain
configuration and conjugate momentum variables respectively, specifically
holonomy -- and electric flux variables. We stress that we use here  
non-standard flux variables not previously considered in the literature.
These non-local functions on the 
classical phase space are, in particular, used to regularize more 
complicated composite operators such as the geometrical operators 
mentioned above. It is already quite remarkable that one can remove the 
regulator {\it without encountering any UV divergencies} ! However,
in order to be convinced that this regulator-free operator really has the 
correct 
classical limit one has to check, for instance, that it has the correct
expectation values with respect to semi-classical states. The question 
arises how such semi-classical states should look like. Now, since
the final regulator-free operator is actually only densely defined (since 
it is usually unbounded) one has to employ states which are semi-classical
and {\it simultaneously belong to a dense subspace of the Hilbert space}.
The states which belong to the domain of definition of the operator 
are labelled by graphs $\gamma$. Given such a graph $\gamma$ one can 
define unambiguously holonomies along its edges as the basic 
configuration operators labelled by $\gamma$, however, {\it the associated
(conjugate) momentum operators are largely ambiguous} in the sense 
that any choice of surfaces which are mutually disjoint and are intersected
by precisely one edge of $\gamma$ 
gives rise to completely
identical Poisson brackets between the canonical variables. 

This then 
leads to the following problem : Suppose we define 
a semi-classical state by requiring that the expectation value of the 
basic holonomy and electric flux operators associated with $\gamma$ take
certain values and satisfy a minimal uncertainty property. Given a point
in the classical phase space those values should be the values of the 
holonomy and flux functions evaluated at that point. However, {\it this 
makes sense only when we specify the surfaces with respect to which we 
calculate the flux}. 

We are thus led to invent a new kind of generalized projective family 
labelled
not only by graphs but also by so-called ``dual'' faces. In particular, we 
wish to do this already at the classical level by introducing a new
kind of generalized projective family of symplectic manifolds. The idea 
behind all of 
this is that these symplectic manifolds enable us to discuss in a clean
way the quantization procedure and its inverse, the process of 
taking the classical limit :
\begin{itemize}
\item[1.] {\it Classical Regularization}\\ 
Suppose we are given a function on the classical phase space $(M,\Omega)$, 
usually
a function $F(m)$ of the connection and the electric field, $m=(A,E)$. 
Here $M$ denotes the set of connections and electric fields respectively
(a differentiable manifold modelled on a Banach space, see below) and
$\Omega$ is a strong symplectic structure on $M$.
As we 
cannot define $\hat{A},\hat{E}$ on our Hilbert space directly as operators, 
we must first find a substitute $F_\gamma(m)$ for $F$ which can be 
written entirely in terms of holonomy and flux variables associated 
with $\gamma$. These variables coordinatize a symplectic manifold
$(M_\gamma,\Omega_\gamma)$. We will say that $F_\gamma(m)$ is a 
substitute for $F(m)$ provided that 1) $F_\gamma$ converges to $F$
pointwise on $M$ as $\gamma\to\infty$ (the graph becomes infinitely fine,
we will specify this limit below) and 2) that the Hamiltonian vector 
field of $F_\gamma$ with respect to $\Omega_\gamma$ converges pointwise 
on $M$ to that of $F$ with respect to $\Omega$.
\item[2.] {\it Regularized Operators}\\
The classical phase spaces $(M_\gamma,\Omega_\gamma)$ turn out to be 
(in)finite direct products of (copies of) cotangent spaces over the gauge 
group $G$ equipped with a non-standard symplectic structure and allow for a 
bona fide quantization by usual geometrical quantization techniques.
By substituting classical variables for operators defined 
on a subspace ${\cal H}_\gamma$ of the Hilbert space and Poisson brackets
with respect to $\Omega_\gamma$ by commutators we obtain an operator 
$\hat{F}_\gamma$ unambiguously defined on ${\cal H}_\gamma$ up to factor
ordering ambiguities. Thus, the phase spaces $M_\gamma$ are much better
suited for the quantization of interesting functions $F$ on $M$ as 
they are automatically finite and we have always control that the
quantization has the correct classical limit on $M_\gamma$. In other 
words, quantization and regularization can be neatly separated as 
individual processes.
\item[3.] {\it Unregularized Operator}\\
It turns out that for a large class of functions $F$ including the ones
of physical interest the family of 
operators $\hat{F}_\gamma$ so obtained provides an operator $\hat{F}$
consistently defined on a dense subspace of the whole Hilbert space in the 
sense that its restriction to ${\cal H}_\gamma$ coincides with 
$\hat{F}_\gamma$. This will be our candidate for a well-defined 
continuum operator.
\item[4.] {\it Classical Limit}\\
In order to study the classical limit of $\hat{F}$ we introduce  
a generalized projective family of semi-classical states 
$\psi^t_{\gamma,m}\in{\cal H}_\gamma$ labelled by 
the graph $\gamma$, a point in $m\in M$ and a classicality parameter
$t\propto \hbar$. We say that $\hat{F}_\gamma$ is a 
quantization of $F_\gamma(m)$ provided that 
$\lim_{t\to 0} <\psi^t_{\gamma,m},\hat{F}_\gamma\psi^t_{\gamma,m}>
=F_\gamma(m)$ for each $m\in M$
and that $\hat{F}$ is a quantization of $F$ provided that
$\lim_{t\to 0}[\lim_{\gamma\to\infty}
<\psi^t_{\gamma,m},\hat{F}_\gamma\psi^t_{\gamma,m}>]=F(m)$ for each 
$m\in M$.\\
\end{itemize}
Theses four steps provide then a closed path of how to go from a classical 
phase space function to an operator and back. As we see, this procedure
requires as a classical cornerstone the analysis of the phase 
spaces $(M_\gamma,\Omega_\gamma)$ which is the subject of the present 
paper. In particular, one must show that these symplectic manifolds
contain a generalized projective sequence that can be identified with 
$(M,\Omega)$.\\ 
\\
The outline of the paper is as follows :
 
In section two we recall a working collection of material from the 
kinematical framework of the theory.

In section three we derive from the symplectic manifold $(M,\Omega)$ for 
gauge theories with compact gauge groups (in any dimension and on any 
(globally hyperbolic) manifold) 
a generalized projective family of (in)finite dimensional symplectic 
manifolds
$(M_\gamma,\Omega_\gamma)$ labelled by graphs $\gamma$ embedded in that 
manifold. We show that the generalized projective 
limit symplectic manifold of a certain generalized projective sequence 
agrees with the 
standard symplectic manifolds 
$(M,\Omega)$ for gauge theories (weighted Sobolev spaces). 
The purpose of doing this is
that the generalized family of symplectic manifolds is much better suited to
quantization than the standard gauge theory phase space as outlined above.

In section four we propose a substitute $G_\gamma$ for an important 
function $G$ on the phase space of any gauge theory, 
namely the Gauss constraint and show that $G_\gamma$ converges to $G$
pointwise on $M$ in the generalized projective limit.

In section five we derive the quantization of $(M_\gamma,\Omega_\gamma)$ and
$G_\gamma$. We show that $\hat{G}_\gamma$ is a consistently defined system
of cylindrical projections of an operator $\hat{G}$ whose constraint algebra
closes without anomalies. Finally we sketch the last step of the 
above programme applied to $\hat{G}$ concerning the classical limit. 
The proof that this step can be completed will be found in 
\cite{46,47,45}.

In section six we complement earlier results obtained by
Ashtekar, Corichi and Zapata \cite{40} :\\ 
These authors considered certain classical functions
$F$ on $M$ and quantized them using $\Omega$ as a starting point.
They obtained operators $\hat{F}$ this way which do not commute 
on certain states $f_\gamma\in {\cal H}_\gamma$ although
the classical functions $F$ have vanishing Poisson brackets (with respect
to $\Omega$) among each other. 
This seeming quantum ``anomaly'' was explained by  
pointing out that the connection and electric field of the theory are
smeared with distributional rather than smooth test functions. If one
uses a smearing with smooth functions then the ``anomaly'' vanishes,
allowing the interpretation that the Poisson brackets of the 
unsmeared fields is non-vanishing, proportional to a distribution with 
support contained in a measurable subset of Lebesgue measure zero 
which is therefore detectable only when smearing with distributional
smearing functions. This interpretation therefore removes the apparent
contradiction. However, then one notices 
that this extended Poisson bracket does not close in an obvious way
(the Jacobi identity is not obeyed in an obvious way). This 
was shown not to be an obstacle to quantization by recalling that it is not 
necessary to base the quantization on Poisson brackets but that one can 
instead base it on the Lie algebra of vector fields on $M$ {\it which
always obey the Jacobi identity and is always closed}.

We show that the non-commutativity of these operators has a natural 
explanation from the point of view of the symplectic manifolds
$(M_\gamma,\Omega_\gamma)$ : \\
1) We observe that we can find, for each of the above choices of $\gamma$, 
functions $F_\gamma\not=F$ which can be considered
as functions on $M_\gamma$ as well. Furthermore, the functions
$F_\gamma$ do have non-vanishing Poisson brackets among each other,
both with respect to $\Omega$ and  with respect to $\Omega_\gamma$
(actually, their brackets with respect to $\Omega_\gamma$ {\it follow}
from those with resepct to $\Omega$).\\ 
2) The quantization of these new 
functions is such that $\hat{F}_\gamma$ and $\hat{F}$ agree on 
${\cal H}_\gamma$.\\
3) The commutator algebra of the $\hat{F}$ on ${\cal H}_\gamma$
is {\it precisely} the one to be expected from the Poisson bracket
structure of the $F_\gamma$.\\
In conclusion, the unexpected non-commutativity observed in \cite{40} can 
be related to a quantization ambiguity. If we insist on a Poisson bracket
-- comutator correspondence principle, however, then one cannot accept 
the $F$ as classical limit of $\hat{F}$ but must instead consider the 
$F_\gamma$.

Finally, in an appendix we write the symplectic structure
$\Omega_\gamma$ for $G=U(1),SU(2)$ in the language of differential forms
which could be useful for future research.

\section{Preliminaries}
\label{s2}

In this section we will recall the main ingredients of the mathematical
formulation of diffeomorphism invariant quantum field theories of 
connections with local degrees of freedom in any dimension and for
any compact gauge group. See \cite{7} and references therein
for more details.\\

Let $G$ be a compact gauge group, $\Sigma$ a $D-$dimensional manifold which
admits a principal $G-$bundle with connection over $\Sigma$.
Let us denote the pull-back to $\Sigma$ of the connection by 
local sections by $A_a^i$
where $a,b,c,..=1,..,D$ denote tensorial indices and $i,j,k,..=1,..,
\dim(G)$ denote indices for the Lie algebra of $G$. We will denote the set
of all smooth connections by $\a$ and endow it with a globally 
defined metric topology of the Sobolev kind
\be \label{2.1}
d_\rho[A,A']:=\sqrt{-\frac{1}{N}\int_\Sigma d^Dx \sqrt{\det(\rho)(x)} 
\mbox{tr}([A_a-A'_a](x)[A_b-A'_b](x))\rho^{ab}(x)}
\ee
where $\mbox{tr}(\tau_i\tau_j)=-N\delta_{ij}$ is our choice of 
normalization for the generators of a Lie algebra $Lie(G)$ of rank $N$ and
our conventions are such
that $[\tau_i,\tau_j]=2f_{ij}\;^k\tau_k$ define the structure constants 
of $Lie(G)$. Here $\rho_{ab}$ is a fiducial metric on $\Sigma$ of 
everywhere Euclidean
signature. In what follows we assume that either $D\not=2$ (
for $D=2$, (\ref{2.1}) depends only on the conformal structure
of $\rho$ and cannot guarantee convergence for arbitrary fall-off 
conditions on the connections) or that $D=2$ and the fields $A$ 
are Lebesgue integrable.  
 
Let $\Gamma^\omega_0$ be the set
of all piecewise analytic, oriented graphs $\gamma$ embedded into $\Sigma$ 
and denote by $E(\gamma)$ and $V(\gamma)$ respectively its sets of oriented
edges $e$ and vertices $v$ respectively. One can extend the framework to 
certain, tame piecewise smooth graphs \cite{32,33} but the description becomes 
more complicated and we refrain from doing this here. More important is
the extension to infinite piecewise analytical graphs $\Gamma^\omega_\sigma$
about which
much will be said in the first reference of \cite{45}. For the purpose of this
paper it will be sufficient to stick to $\Gamma^\omega_0$ which is 
sufficient, e.g. if $\Sigma$ is compact. All the properties that are derived
here for $\Gamma^\omega_0$ readily extend to $\Gamma^\omega_\sigma$ as one 
can easily check. 

We denote by $h_e(A)$ the holonomy
of $A$ along $e$ and say that a function $f$ on $\a$ is cylindrical with 
respect to $\gamma$ if there exists a function $f_\gamma$ on 
$G^{|E(\gamma)|}$ such that $f=p_\gamma^\ast f_\gamma=f\circ p_\gamma$ 
where $p_\gamma(A)=\{h_e(A)\}_{e\in E(\gamma)}$. The set of such 
functions is denoted by $\Phi_\gamma$. Holonomies are invariant under
reparameterizations of the edge and in this article we take
edges always to be analytic diffeomorphisms between $[0,1]$ and a
one-dimensional submanifold of $\Sigma$. Gauge transformations are functions
$g:\;\Sigma\mapsto G;\;x\mapsto g(x)$ and they act on
holonomies as $h_e\mapsto g(e(0))h_e g(e(1))^{-1}$. 

A particularly useful set of cylindrical functions are the so-called 
spin-netwok functions \cite{34,35,9}. A spin-network function is 
labelled by a graph $\gamma$, a set of irreducible representations 
$\vec{\pi}=\{\pi_e\}_{e\in E(\gamma)}$ (choose from each equivalence 
class of equivalent
representations once and for all a fixed representant), one for each 
edge of $\gamma$, and a set $\vec{c}=\{c_v\}_{v\in V(\gamma)}$ of
contraction matrices, one for each vertex of $\gamma$, which 
contract the indices of the tensor product 
$\otimes_{e\in E(\gamma)} \pi_e(h_e)$ in such a way that the resulting
function is gauge invariant. We denote spin-network functions as
$T_I$ where $I=\{\gamma,\vec{\pi},\vec{c}\}$ is a compound label.
One can show that these functions are linearly independent.

The set of finite linear combinations of spin-network functions forms an 
Abelian $^\ast$ algebra $\cal B$
of functions on $\a$. By completing it with respect to the sup-norm 
topology it 
becomes an Abelian C$^\ast$ algebra (here the compactness of $G$ is 
crucial). The spectrum $\ab$ of this algebra, 
that is, the set of all algebraic homomorphisms ${\cal B}\mapsto\Co$
is called the quantum configuration space. This space is equipped with
the Gel'fand topology, that is, the space of continuous functions
$C^0(\ab)$
on $\ab$ is given by the Gel'fand transforms of elements of $\cal B$.
Recall that the Gel'fand transform is given by $\tilde{f}(\bar{A}):=
\bar{A}(f)\;\forall \bar{A}\in \ab$. It is easy to see that $\ab$ with 
this topology is a compact Hausdorff space. Obviously, the elements of
$\a$ are contained in $\ab$ and one can show that $\a$ is even dense
\cite{36}. Generic elements of $\ab$ are, however, distributional.

The idea is now to construct a Hilbert space consisting of square
integrable functions on $\ab$ with respect to some measure $\mu$. Recall 
that one can define a measure on a locally compact Hausdorff space 
by prescribing a positive linear functional $\chi_\mu$ on the space 
of continuous functions thereon. The particular measure
we choose is given by $\chi_{\mu_0}(\tilde{T}_I)=1$ if $I=\{\{p\},
\vec{0},\vec{1}\}$ and $\chi_{\mu_0}(\tilde{T}_I)=0$ otherwise. Here
$p$ is any point in $\Sigma$, $0$ denotes the 
trivial representation and $1$ the trivial contraction matrix. In other 
words, (Gel'fand transforms of) spin-network functions play the same role 
for $\mu_0$ as 
Wick-polynomials do for Gaussian measures and like those they form
an orthonormal basis in the Hilbert space ${\cal H}:=L_2(\ab,d\mu_0)$ 
obtained by completing their finite linear span $\Phi$.\\
An equivalent definition of $\ab,\mu_0$ is as follows :\\ 
$\ab$ is in one to one correspondence, via the surjective map $H$ defined 
below, with the set $\ab':=\mbox{Hom}({\cal X},G)$
of homomorphisms from the groupoid $\cal X$ of composable, holonomically
independent, analytical paths
into the gauge group. The correspondence is explicitly given by
$\ab\ni\bar{A}\mapsto H_{\bar{A}}\in\mbox{Hom}({\cal X},G)$
where ${\cal X}\ni e\mapsto H_{\bar{A}}(e):=\bar{A}(h_e)=
\tilde{h}_e(\bar{A})\in G$ and $\tilde{h}_e$ is the Gel'fand transform
of the function $\a\ni A\mapsto h_e(A)\in G$. Consider now the restriction
of $\cal X$ to ${\cal X}_\gamma$, the groupoid of composable edges of  
the graph $\gamma$. One can then show that the projective limit of the 
corresponding {\it cylindrical sets} 
$\ab'_\gamma:=\mbox{Hom}({\cal X}_\gamma,G)$ coincides with $\ab'$.
Moreover, we have $\{\{H(e)\}_{e\in E(\gamma)};\;H\in\ab'_\gamma\}=
\{\{H_{\bar{A}}(e)\}_{e\in E(\gamma)};\;\bar{A}\in\ab\}=
G^{|E(\gamma)|}$.
Let now $f\in{\cal B}$ be a function cylindrical over $\gamma$ then 
$$
\chi_{\mu_0}(\tilde{f})=\int_{\ab} d\mu_0(\bar{A}) \tilde{f}(\bar{A})
=\int_{G^{|E(\gamma)|}} \otimes_{e\in E(\gamma)} d\mu_H(h_e)
f_\gamma(\{h_e\}_{e\in E(\gamma)})
$$
where $\mu_H$ is the Haar measure on $G$.
As usual, $\a$ turns out to be contained in a measurable subset of 
$\ab$ which has measure zero with respect to $\mu_0$.

This concludes the definition of the kinematical Hilbert space $\cal H$,
of the quantum configuration space $\ab$ and of the classical
configuration space. What about the classical and quantum phase space ? 
This question has actually so far not been analysed satisfactorily
in the literature, partial results are scattered over a number of papers.
We therefore begin the next section with this issue.

\section{Symplectic Manifolds Labelled by Graphs}
\label{s3}

\subsection{Standard Continuum Symplectic Structures}
\label{s3.1}

Let us first recall the usual infinite dimensional symplectic geometry that 
underlies gauge theories.\\

Let $F^a_i$ be a Lie algebra valued vector density test field of weight one 
and let $f_a^i$ be a Lie algebra valued covector test field. Let,
as before $A_a^i$ be a the pull-back of a connection to $\Sigma$ and consider
a vector bundle of electric fields, that is, of Lie algebra valued 
vector densities of weight one whose bundle projection to $\Sigma$ we denote 
by $E^a_i$. We consider the smeared quantities
\be \label{3.1}
F(A):=\int_\Sigma d^Dx F^a_i A_a^i\mbox{ and } 
E(f):=\int_\Sigma d^Dx E^a_i f_a^i 
\ee
While both are diffeomorphism covarinat it is only the latter which is 
gauge covariant, one reason to consider the singular smearing functions
discussed below.
The choice of the space of pairs of test fields $(F,f)\in{\cal S}$ 
depends on the boundary conditions on
the space of connections and electric fields which in turn depends on the 
topology of $\Sigma$ and will not be specified in what follows. 

Consider the set $M$
of all pairs of smooth functions $(A,E)$ on $\Sigma$ such that (\ref{3.1}) is 
well defined for any $(F,f)\in {\cal S}$. 
We wish to endow it with a manifold structure and a symplectic 
structure, that is, we wish to turn it into an infinite dimensional 
symplectic manifold.

We define a topology on $M$ through the metric :
\ba \label{3.2}
&& d_{\rho,\sigma}[(A,E),(A',E')]
\nonumber\\
&:=&\sqrt{-\frac{1}{N}\int_\Sigma d^Dx 
[\sqrt{\det(\rho)} \rho^{ab} \mbox{tr}([A_a-A'_a][A_b-A'_b])+
\frac{\sigma_{ab} \mbox{tr}([E^a-E^{a\prime}][E^b-E^{b\prime}])}
{\sqrt{\det(\sigma)}}]} 
\ea
where $\rho_{ab},\sigma_{ab}$ are again fiducial metrics on $\Sigma$ of 
everywhere Euclidean signature. Their fall-off behaviour has to be suited
to the boundary conditions of the fields $A,E$ at spatial infinity.
Notice that the metric (\ref{3.2}) is gauge invariant (and thus globally 
defined) and diffeomorphism covariant and that 
$d_{\rho,\sigma}[(A,E),(A',E')]=
d_\rho[A,A']+d_\sigma[E,E']$ (recall (\ref{2.1})).

Now, while the space of electric fields in Yang-Mills theory is a vector 
space, the 
space of connections is only an affine space. However, as we have also 
applications in general relativity with asymptotically Minkowskian
boundary conditions in mind, also the space of electric fields 
will in general not be a vector space.
Thus, in order to induce a norm
from (\ref{3.2}) we proceed as follows : Consider an atlas of $M$
consisting only of $M$ itself and 
choose a fiducial background connection and electric field $A^{(0)}, 
E^{(0)}$ (for instance $A^{(0)}=0$). We define the global chart 
\be \label{3.3}
\varphi\; :\; M\mapsto {\cal E};\; (A,E)\mapsto (A-A^{(0)},E-E^{(0)})
\ee
of $M$ onto the vector space of pairs $(A-A^{(0)},E-E^{(0)})$. Obviously,
$\varphi$ is a bijection. We topologize $\cal E$ in the norm
\be \label{3.4}
||(A-A^{(0)},E-E^{(0)})||_{\rho\sigma}:=
\sqrt{d_{\rho\sigma}[(A,E),(A^{(0)},E^{(0)})]}
\ee
The norm (\ref{4.4}) is of course no longer gauge and diffeomorphism 
covariant since the fields $A^{(0)},E^{(0)}$ do not transform, they are
backgrond fields. We need it, however, only in order to encode the 
fall-off behaviour of the fields which are independent of gauge -- and
diffeomorphism covariance.
 
Notice that the metric induced by this norm coincides with (\ref{3.2}).
In the terminology of weighted Sobolev spaces the completion of $\cal E$ in 
the norm (\ref{3.4}) is called the Sobolev space
$H^2_{0,\rho}\times H^2_{0,\sigma^{-1}}$, see e.g. \cite{37}. We will call
the completed space $\cal E$ again and its image under 
$\varphi^{-1}$, $M$ again (the dependence of $\varphi$ on 
$(A^{(0)},E^{(0)})$ will be suppressed). Thus, $\cal E$ is a 
normed, complete
vector space, that is, a Banach space, in fact it is even a Hilbert space.
Moreover, we have modelled $M$ on the Banach space $\cal E$, that is, 
$M$ acquires the structure of a (so far only topological) Banach manifold.
However, since $M$ can be covered by a single chart and the identity map
on $\cal E$ is certainly $C^\infty$, $M$ is actually a smooth manifold.
The advantage of modelling $M$ on a Banach manifold is that one can
take over almost all the pleasant properties from the finite dimensional 
case to the infinite dimensional one (in particular, the inverse function 
theorem). 

Next we study differential geometry on $M$ with the standard techniques 
of calculus on infinite dimensional manifolds (see e.g. \cite{38,39}).
We will not repeat all the technicalities of the definitions involved,
the interested reader is referred to the literature quoted.
\begin{itemize}
\item[i)] 
A function $f:\; M\mapsto \Co$ on $M$ is said to be differentiable 
at $m$
if $g:=f\circ\varphi^{-1}:\;{\cal E}\mapsto \Co$ is differentiable
at $u=\varphi(m)$,
that is, there exist {\it bounded} linear operators $Dg_u, Rg_u:\;\cal 
E\mapsto\Co$ such that 
\be \label{3.5}
g(u+v)-g(u)=(Dg_u)\cdot v+(Rg_u)\cdot v \mbox{ where }
\lim_{||v||\to 0} \frac{|(Rg_u)\cdot v|}{||v||}=0\;.
\ee
$Df_m:=Dg_u$ is called the functional derivative of $f$ at $m$ (notice
that we identify, as usual, the tangent space of $M$ at $m$ with 
$\cal E$). The definition extends in an obvious way to the case where
$\Co$ is replaced by another Banach manifold. The equivalence class
of functions differentiable at $m$ is called the germ $G(m)$ at $m$.
Here two functions are said to be equivalent provided they coincide in 
a neighbourhood containing $m$.
\item[ii)] In general, a tangent vector $v_m$ at $m\in M$ is an equivalence 
class of triples 
$(U,\varphi,v_m)$ where $(U,\varphi)$ is a chart of the atlas of $M$ 
containing $m$ and $v_m\in {\cal E}$. 
Two triples are said to be equivalent provided that
$v'_m=D(\varphi'\circ\varphi^{-1})_{\varphi(m)}\cdot v_m$. In our case 
we have only one chart and equivalence becomes trivial. Tangent vectors
at $m$ can be considererd as derivatives on the germ $G(m)$ by defining
\be \label{3.6}
v_m(f):=(Df_m)\cdot v_m=(D(f\circ\varphi^{-1})_{\varphi(m)})\cdot v_m
\ee
Notice that the definition depends only on the equivalence class and not
on the representant. The set of vectors tangent at $m$ defines the tangent
space $T_m(M)$ of $M$ at $m$.
\item[iii)] The cotangent space $T'_m(M)$ is the topological dual
of $T_m(M)$, that is, the set of {\it continuous} linear functionals on 
$T_m(M)$. It is obviously isomorphic with ${\cal E}'$, the topological 
dual of $\cal E$. Since our model space $\cal E$ is reflexive (it is a 
Hilbert space) we can naturally identify tangent and cotangent space
(by the Riesz lemma) which also makes the definition of contravariant 
tensors less ambiguous. We will, however, not need them for what follows. 
Similarly, one defines the space of
$p-$covariant tensors at $m\in M$ as the space of {\it continuous} 
$p-$linear forms on the $p-$fold tensor product of $T_m(M)$.
\item[iv)] So far the fact that $\cal E$ is a Banach manifold was not very
crucial. But
while the tangent bundle $T(M)=\cup_{m\in M} T_m(M)$ carries a 
natural manifold structure modelled on ${\cal E}\times {\cal E}$
for a general Fr\'echet space (or even locally convex space) $\cal E$
the cotangent bundle $T'(M)=\cup_{m\in M} T'_m(M)$ 
carries a manifold structure only when $\cal E$ is a Banach space as one 
needs the inverse function theorem to show that each chart is not only a 
differentiable bijection but that also its inverse is differentiable.
In our case again there is no problem. We define differentiable vector 
fields and $p-$covariant tensor fields as cross sections of the 
corresponding fibre bundles.
\item[v)] A differential form of degree $p$ on $M$ or $p-$form is a cross 
section 
of the fibre bundle of completely skew continuous $p-$linear forms.
Exterior product, pull-back, exterior differential, interior product
with vector fields and Lie derivatives are defined as in the finite 
dimensional case.
\end{itemize}
\begin{Definition} \label{def3.1}
Let $M$ be a differentiable manifold modelled on a Banach space $\cal E$.
A weak respectively strong symplectic structure $\Omega$ on $M$ is a closed 
2-form such that for all $m\in M$ the map
\be \label{3.7}
\Omega_m:\; T_m(M)\mapsto T'_m(M);\; v_m\to \Omega(v_m,.)
\ee
is an injection respectively a bijection.
\end{Definition}
Strong symplectic structures are more useful because weak symplectic 
structures do not allow us to define Hamiltonian vector fields
through the definition $DL+i_{\chi_L}\Omega=0$ for differentiable 
$L$ on $M$ and Poisson brackets through
$\{f,g\}:=\Omega(\chi_f,\chi_g)$. Thus we define finally a strong 
symplectic structure for our case by
\be \label{3.8}
\Omega((f,F),(f',F')):=\int_\Sigma d^Dx [F^a_i f^{i\prime}_a
-F^{a\prime}_i f_a^i](x)
\ee
for any $(f,F),(f',F')\in {\cal E}$. 
To see that $\Omega$ is a strong symplectic structure
we observe first that the integral kernel of $\Omega$ is constant so that
$\Omega$ is clearly exact, so, in particular, closed. Next, let
$\theta\in {\cal E}'\equiv {\cal E}$. To show that $\Omega$ is a bijection 
it suffices 
to show that it is a surjection (injectivity follows trivially from 
linearity). We must find $(f,F)\in {\cal E}$ so that $\theta(.)=
\Omega((f,F),.)$. Now by the Riesz lemma there exists 
$(f_\theta,F_\theta)\in {\cal E}$ such that $\theta(.)=
<(f_\theta,F_\theta),.>$ where $<.,.>$ is the inner product induced 
by (\ref{3.4}). Comparing (\ref{3.4}) and (\ref{3.8}) we see that we have 
achieved our goal provided that the functions 
\be \label{3.9}
F^a_i:=\rho^{ab}\sqrt{\det(\rho)}f_{b\theta}^i,\;
f_a^i:=-\frac{\sigma_{ab}}{\sqrt{\det(\rho)}}F^b_{i\theta}
\ee
are elements of $\cal E$. Inserting the definitions we see that this will 
be the case provided that the functions 
$\rho^{cd}\sigma_{ca}\sigma_{db}/\sqrt{\det(\rho)}$ and
$\det(\rho)\sigma_{cd}\rho^{ca}\rho^{db}/\sqrt{\det(\sigma)}$ 
respectively fall off at least as 
$\sigma_{ab}/\sqrt{\det(\sigma)}$ and 
$\rho^{ab}\sqrt{\det(\rho)}$ respectively. In physical 
applications these metrics are usually chosen to be of the form
$1+O(1/r)$ where $r$ is an asymptotical radius function so that these
conditions are certainly satisfied.
Therefore, $(f,F)\in {\cal E}$ and our small lemma is established.

Let us compute the Hamiltonian vector field of a function $L$ on our $M$.
By definition for all $(f,F)\in {\cal E}$ we have at $m=(A,E)$
\be \label{3.10}
DL_m\cdot(f,F)=\int_\Sigma d^Dx [(DL_m)^a_i f_a^i+(DL_m)_a^i F^a_i]
=-\int_\Sigma d^Dx [(\chi_{Lm})^a_i f_a^i-(\chi_{Lm})_a^i F^a_i
\ee
thus $(\chi_L)^a_i=-(DL)^a_i$ and $(\chi_L)_a^i=(DL)_a^i$. Obviously,
this defines a bounded operator on $\cal E$ if and only if $L$ is
differentiable. Finally, the Poisson bracket is given by
\be \label{3.11}
\{L,L'\}_m=\Omega(\chi_L,\chi_{L'})=\int_\Sigma d^Dx
[(DL_m)_a^i (DL'_m)^a_i-(DL_m)^a_i (DL'_m)_a^i]
\ee
It is easy to see that $\Omega$ has the symplectic potential $\Theta$,
a one-form on $M$, defined by
\be \label{3.12}
\Theta_m((f,F))=\int_\Sigma d^Dx E^a_i f_a^i
\ee
since 
$$
D\Theta_m((f,F),(f',F'))
:=(D(\Theta_m)\cdot(f,F))\cdot (f',F')
-(D(\Theta_m)\cdot(f',F'))\cdot (f,F)
$$
and $DE^a_i(x)_m\cdot (f,F)=F^a_i(x)$ as follows from the definition.

Coming back to the choice of $\cal S$, it will in general be a subspace 
of $\cal E$ so that (\ref{3.1}) still converges. We can now compute the 
Poisson brackets between the functions $F(A),E(f)$ on $M$ and find 
\be \label{3.13}
\{E(f),E(f')\}=\{F(A),F'(A)\}=0,\;\{E(f),A(F)\}=F(f)
\ee
Remark :\\
In physicists notation one often writes $(DL_m)_a^i(x):=
\frac{\delta L}{\delta A_a^i(x)}$ etc. and one writes the symplectic 
structure as
$\Omega=\int d^Dx \;DE^a_i(x)\wedge DA_a^i(x)$.

\subsection{New Symplectic Structure}
\label{s3.2}

As outlined in the introduction we wish to define symplectic manifolds 
$(M_\gamma,\Omega_\gamma)$ for every $\gamma\in\Gamma$ which ``in some sense
come from $(M,\Omega)$''. In order to do this we need besides graphs new
extended objects associated with them. This is the topic of the following
subsubsection. The class of graphs $\gamma$ that we have in mind consists 
of the 
set $\Gamma^\omega_\sigma$ of piecewise analytic, $\sigma$-finite graphs.
These are graphs with an at most countable number of edges and such that
for every compact subset $U$ of the locally compact manifold $\Sigma$
the restriction $U\cap\gamma$ is a piecewise analytic, finite graph
in $\Gamma^\omega_0$. The precise definition of these graphs and their 
properties as well as the extension of the quantum kinematical framework to 
them needs the framework of the infinite tensor product of Hilbert spaces
and will therefore be postponed to the first reference of \cite{45}.
For the rest of this paper one may without doing any mistake think of 
$\gamma$ as an element of $\Gamma^\omega_0$.

\subsubsection{Dual Polyhedral Decompositions}
\label{s3.2.0}

\begin{Definition} \label{def3.2}
A polyhedral decomposition $P$ of $\Sigma$ is a subdivision of $\Sigma$
into closed compact regions $\Delta$, that is $\Sigma=\cup_{\Delta\in P} 
\Delta$, 
each of which is diffeomorphic to a polyhedron in flat space
and intersects any other polyhedron only in 
the set of points of their common boundary (of codimension at least one).
\end{Definition}
Note that we allow
for decompositions with a countably infinite number of polyhedra in case that
$\Sigma$ is not compact. We also do not insist that the decomposition be 
convex as this would require the choice of a background metric 
(rather of a diffeomorphism invariance class because convexity is 
defined by geodesity of curves which is a diffeomorphism invariant notion).
\begin{Definition} \label{def3.3}
Let $P$ be a polyhedronal decomposition of $\Sigma$, pick any
polyhedron $\Delta\in P$ and consider its boundary $\partial\Delta$.

i) A face $S$ of $\Delta$ (a maximal, connected, analytic subset of 
its boundary of codimension one) is 
called a ``standard face'' provided that\\
1) $S$ is a submanifold of $\Sigma$ of codimension one, diffeomorphic
to a standard cube $[-1,1]^{D-1}$ in $\Rl^{D-1}$.\\
2) $S$ has no boundary, $\partial S=\emptyset$, i.e. it is open.\\
3) $S$ is contained in the domain of a chart of $\Sigma$.\\
4) $S$ is maximal, that is, there does not exist any 
$S'\subset\partial\Delta$ properly containing $S$ which satisfies 1), 2) 
and 3).

ii) As $S$ is an open submanifold of $\Sigma$ of codimension one and for some
$\Delta\in P,\; S\subset\partial\Delta$ is contained in the domain of a 
chart $(U,\varphi)$ of an atlas of $\Sigma$ there exists an open subset
$V\subset U$ containing $S$, divided by $S$ into two halves and a 
diffeomorphism $\varphi'$
that maps $V,S$ respectively to $\Rl^D$ and the hypersurface $x^D=0$ 
respectively. An orientation of $S$ is given by a choice of which of the 
half spaces or ``sides''  given by the set of points satisfying $x^D>0$ or 
$x^D<0$ we call ``up'' or ``down''.

iii) A polyhedronal decomposition of $\Sigma$ is said to be oriented 
provided the collection of all its standard faces have been oriented.
\end{Definition}
This definition just formalizes the intuitive idea of a face of a polyhedron
with a regular shape.
Notice that a face is always shared by precisely two polyhedra of the 
decomposition.
In D=1,2,3 respectively simplicial polyhedra are given by closed lines,
triangles and tetrahedra respectively and their faces are open points,
lines and triangles respectively. The notion of an orientation is clearly 
independent of the chart employed. Also, faces are always orientable even 
if $\Sigma$ is not orientable and even if $\Sigma$ is orientable the 
orientation of a face can be opposite to the orientation induced from 
$\Sigma$ on $S$ (as a submanifold).
\begin{Definition} \label{def3.4}
Given a graph $\gamma\in\Gamma$ we say that an oriented, polyhedral 
decomposition $P$ of $\Sigma$ is {\it dual} to $\gamma$ provided that :\\
1) Given an edge $e$ of $\gamma$ there exists precisely one standard 
face $S_e$ from the collection of all faces of all the polyhedra of the 
decomposition which is intersected by $e$.\\
2) The edge $e$ intersects $S_e$ transversally, that is, a)
$\{p_e\}:=e\cap S_e$ consists of a single point which is an interior 
point of $e$ (it is necessarily an interior point of $S_e$ since $S_e$ is 
open)
and b) there exists an open neighbourhood $V$ of $p_e$ and a diffeomorphism
which maps $V$ to $\Rl^D$, $p_e$ to the origin of $\Rl^D$, $S_e\cap V$
to the plane $x^D=0$ and $e\cap V$ to the line $x^1=..=x^{D-1}=0$.\\
3) The orientations of $S_e$ and $e$ agree, that is, if under the 
diffeomorphism outlined in 2) the edge points into the half space
$x^D>0$ or $x^D<0$ respectively then that half space corresponds to the
``up'' side of $S_e$. \\
4) The decomposition is irreducible, that is, one cannot reduce the 
number of polyhedra of the decomposition (by removing faces) without 
destroying at least one of the properties 1)-3).
\end{Definition}
Dual decompositions certainly exist in any dimension : A first example is 
given by a cubic lattice in any dimension (every vertex is 2D-valent)
where we assume that $\Sigma$ has no boundary (periodic boundary 
conditions).
The dual lattice is unique up to diffeomorphisms and corresponds to 
D-cubes around every vertex. A second example (again with periodic 
boundary conditions) is given by a simplicial 
lattice (every vertex is (D+1)-valent) and corresponds to a simplicial 
decomposition with D-simplices around every vertex. Again the dual 
lattice is uniquely determined up to diffeomorphisms. Whether or 
not this uniqueness is a general feature of dual decompositions is 
not clear but we have the following.
\begin{Theorem} \label{th3.1}
Given a graph $\gamma\in \Gamma$, a dual, oriented, polyhedral
decomposition $P_\gamma$ of it exists and it is unique up to 
diffeomorphisms and up to the number of possibilities to obey
the Euler (or Dehn-Sommerfeld in $D\not=3$) relation for the various 
polyhedra under the reduction process.
\end{Theorem}
Proof of Theorem \ref{th3.1} :\\
{\it Existence} :\\
Given a graph $\gamma$ consider for each vertex $v$ a closed neighbourhood 
$U_v$ of $v$ such that the various $U_v$ are mutually disjoint to begin with.
Moreover, we can choose them to be closed balls such that their boundaries 
have the topology of the sphere.
Now distort the $U_v$ along the edges $e$ incident at
$v$ without changing their topology until for any edge $e$ with end points 
$v,v'$ the $U_v,U_{v'}$ intersect in precisely one point $p_e$ which
is obviously an interior point of $e$, otherwise they are mutually
non-intersecting. Now blow up the $U_v$ even more,
keeping the $p_e$ fixed until each $U_v$ looks like a solid ball
with a spherical boundary except for $n(v)$ cusps $S_e$ of the topology of 
cubes $[-1,1]^{D-1}$ with the $p_e$ as an interior point. 
$S_e$ is the only set in which $U_v,U_{v'},\;\{v,v'\}=\partial e$ intersect.
By shifting the $p_e$ and varying the size of the
cusps we can achieve that the $S_e$ are mutually non-intersecting, 
contained in the domain of a chart and intersect $e$ transversally.
We can therefore equip them with an orientation that agrees with that of 
$e$. Altogether, the $U_v$ have now mutated to polyhedrons
with $n(v)$ open, smooth standard faces $S_e,\; v\in\partial e$ and the 
additional
closed connected component of its boundary consisting of 
$S_v:=\partial U_v-\cup_{v\in\partial e} S_e$ . Finally, consider the
polyhedronal decomposition of $\Sigma$ consisting of the $U_v$ and 
the remainder $\Sigma-\cup_{v} U_v$. This decomposition satisfies all the 
properties of a dual decomposition except, possibly, for the irreducibility
requirement. To satisfy it, we remove the faces $S_v$ by letting
the $S_e$ share their boundaries, as long as compatible with the 
Euler relation \cite{39a} between the number of connected components of all 
possible subcomplexes of a polyhedron. (For instance, in D=3 
the relation is given by $f=k-e+\chi$ where $f,k,e$ denotes the number of
faces, links and corners of a polyhedron and $\chi$ is essentially the 
Euler characteristic of the manifold that the polyhedron
triangulates). That this is 
always possible follows by the lemma of choice (or Zorn's lemma).\\
{\it Uniqueness} :\\
The constructive proof just given just has just fixed the topology of the 
final dual polyhedronal decomposition reached and is therefore unique
at most up to diffemorphisms. Morover, the number of possible irreducible
decompositions reachable obviously equals the number of possible solutions 
to the just mentioned Euler relation.\\
$\Box$\\
\\
The non-uniqueness of $P_\gamma$ does not affect us because we will
use only the $S_e$ which are determined up to diffeomorphisms.

Notice that on the other hand, if we are given an oriented cellular 
decomposition of $\Sigma$ into polyhedra with open faces, there is, up to 
diffeomorphism equivalence, only one graph such that the decomposition is 
dual to it. This follows easily from the fact that there is no choice but 
choosing an interior point of each polyhedron and connecting common 
faces between polyhedra with transversal edges running between the 
corresponding interior points with the obvious orientation. 

From now on we pick for each graph $\gamma$ an oriented, 
polyhedral, dual decomposition 
$P_\gamma$. By theorem \ref{th3.1} we can do this in such a way that, if 
$\gamma\not=\gamma'$ are diffeomorphic, then $P_\gamma,P_{\gamma'}$ are 
diffeomorphic. Notice, however, that it is not possible to require that
for any diffeomorphism $\varphi$ we have $P_{\varphi(\gamma)}=
\varphi(P_\gamma)$ since there are many diffeomorphisms, say
$\varphi_1,\varphi_2$ such that
$\varphi_1(\gamma)=\varphi_2(\gamma)$ but
$\varphi_1(P_\gamma)\not=\varphi_2(P_\gamma)$.

Also, a word of caution is appropriate with respect to 
refinements : The set $\Gamma$ is partially ordered by inclusion and for 
each pair $\gamma,\gamma'$ there exists a bigger (refined) graph 
$\tilde{\gamma}$ containing both of 
them, for instance the graph $\gamma\cup\gamma'$. However, in general 
it will not be true that there exists a refined graph such that
$P_\gamma,P_{\gamma'}$ are both contained in $P_{\tilde{\gamma}}$. This 
can happen only if the graphs under consideration have a high degree of 
symmetry as e.g. cubic lattices as one can show.

\subsubsection{The Graph Phase Space From the Continuum Phase Space}
\label{s3.2.1}

We are now ready to derive a symplectic manifold $(M_\gamma,\Omega_\gamma)$
for each $\gamma\in \Gamma$ from the standard symplectic manifold
$(M,\Omega)$ considered in the beginning of this section.
\begin{Definition} \label{def3.5} ~\\
i) Let $S_0$ be the interior of the subset $[-1,1]^{D-1}\subset \Rl^{D-1}$
in the $x^D=0$ plane with normal orientation into the $x^D>0$ direction.
Let $p_0=0$ be the origin in $\Rl^D$ and $e_0$ be the interval $[-1/2,1/2]$
of the $x^D$-axis. Let $x_0\in S_0$ and define $\rho_0(x_0)$ to be the 
straight line within $S_0$ connecting $p_0$ and $x_0$.\\
ii) Given a graph $\gamma$ and a dual polyhedronal decomposition
$P_\gamma$, we call a collection of paths $\Pi_\gamma:=
\{\rho_e(x)\subset S_e;x\in S_e\}_{e\in E(\gamma)}$ adapted to 
$\gamma,P_\gamma$ provided there exists for each $e\in E(\gamma)$
an analytic diffeomorphism $\varphi_e:\;\Rl^D\mapsto \Sigma$
such that 
\be \label{3.13a}
(e,S_e,p_e,x,\rho_e(x))
=(\varphi_e(e_0),\varphi_e(S_0),\varphi_e(p_0),\varphi_e(x_0),
\varphi_e(\rho_0(x_0))
\ee
We will denote the set of triples $(\gamma,P_\gamma,\Pi_\gamma)$ 
by ${\cal L}_*$ or $\cal L$ where $*=0,\sigma$ depends on the class
of the graphs that we allow. The elements 
$l\in{\cal L}$ will be called structured 
graphs. We also use the notation
$l=(\gamma(l),P_\gamma(l),
\Pi_\gamma(l))$.\\
iii) Given a structured graph $l$, let w.l.g. $p_e=e(1/2)$.\\
Then we define the following function on $(M,\Omega)$
\be \label{3.14}
P^e_i(A,E):=-\frac{1}{N}
\mbox{tr}(\tau_i h_e(0,1/2)[\int_{S_e} h_{\rho_e(x)} \ast E(x) 
h_{\rho_e(x)}^{-1}] h_e(0,1/2)^{-1})
\ee
where $h_e(s,t)$ denotes the holonomy of $A$ along $e$ between the 
parameter values $s<t$, $\ast$ denotes the Hodge dual, that is,
$\ast E$ is a $(D-1)-$form on $\Sigma$and $E^a:=E^a_i\tau_i$.
\end{Definition}
Notice that in contrast to similar variables used earlier in the literature
\cite{40} the function $P^e_i$ is {\it gauge covariant}. Namely, under gauge 
transformations it transforms as $P^e\mapsto g(e(0)) P^e g(e(0))^{-1}$,
the price to pay being that $P^e$ depends on both $A$ and $E$ and not 
only on $E$. As we will see shortly, this is actually an advantage.
Of course, the notation (\ref{3.14}) is abusing as $P^e$ not only depends on 
$e$ but actually on $S_e,\rho_e(x),x\in S_e$. In the sequel, unless
confusion can arise we will continue abusing notation and write 
$\gamma$ instead of $l$.

The problem with the functions $h_e(A)$ and $P^e_i(A,E)$ on $M$ is that 
they are not differentiable on $M$, that is, $Dh_e, DP^e_i$ are nowhere  
bounded operators on $\cal E$ as one can easily see. The reason for this is,
of course, that these are functions on $M$ which are not properly smeared 
with functions from $\cal S$, rather they are smeared with distributional
test functions with support on $e$ or $S_e$ respectively. Nevertheless
one would like to base the quantization of the theory on these functions 
as basic variables because of their gauge and diffeomorphism covariance.
Indeed, under diffeomorphisms the structured graph $l$
is simply replaced by 
\be \label{3.14a}
\varphi^{-1}(l)=(\varphi^{-1}(e),
\varphi^{-1}(S_e),\{\varphi^{-1}(\rho_e(x));\; x\in S_e\})_{e\in E(\gamma)}
\ee
which is a structured graph again and in this sense 
$h_e\mapsto h_{\varphi^{-1}(e)},P^e_i\mapsto P^{\varphi^{-1}(e)}_i$. 
Furthermore, their quantizations  
are properly represented on the Hilbert space described in section \ref{s2}
as we will see. The fact that the smearing dimensions of $h_e$ and $P^e_i$
add to $D$ raises some hope that one can still derive a bona fide Poisson
algebra among these variables. We therefore define
\begin{Definition} \label{def3.6}
The set of pairs $(h_e(A),P^e_i(A,E))_{e\in E(\gamma)}$ as $(A,E)$ varies 
over $M$ will be denoted by $\bar{M}_{\gamma|M}$. We also define 
$\bar{M}_\gamma=(G\times Lie(G))^{|E(\gamma)|}$.
\end{Definition}
It is easy to see that $\bar{M}_{\gamma|M}$ is generically a proper subset 
of $\bar{M}_\gamma$. Indeed, since the edges $e$ 
are mutually disjoint among each other except for the vertices we can 
find a smooth connection 
with support in disjoint open neighbourhoods $U_e$, one for each $e$,
such that $e\cap U_e$ is an open segment of $e$. The holonomy along those
segments can be given independent values since we can vary the 
behaviour of $A$ in each $U_e$ independently and arbitrarily without 
destroying smoothness. Similar considerations hold for the momenta $P^e_i$.
The range of these values is, however, constrained by the boundary conditions
imposed by the fact that $(A,E)$ are points in a classical phase space 
subject to fall-off conditions. The bar in the notation $\bar{M}_\gamma$
accounts for the fact that the points of $\bar{M}_\gamma$ do not satisfy such
regularity assumptions similar as in the case of $\ab_\gamma$.

We equip a subset $M_\gamma$ of $\bar{M}_\gamma$ with the following natural 
topology : Let 
$(u_i,\phi_i)_{i\in I}$ be an atlas of $G$ where $I$ is a finite index set
(always possible since $G$ is compact). For instance the $u_i$ could
be preimages of open sets in $\Rl^{\dim(G)}$ under the exponential map
which is locally a diffeomorphism between $G$ and $Lie(G)$. Since 
$Lie(G)$ is 
topologically trivial we can construct an atlas of $G\times Lie(G)$ by
$(U_i=u_i\times Lie(G),\Phi_i=\phi_i\times \mbox{id})$ where id denotes 
the identity map of $\Rl^{\dim(G)}$. Then $M_\gamma$ can be given the 
differentiable structure defined by the atlas 
\be \label{3.15}
(\times_{e\in E(\gamma)} U_{i_e},\times_{e\in E(\gamma)} 
\Phi_{i_e})_{i_e\in I}
\ee
which displays $M_\gamma$ as a Banach manifold modelled on 
${\cal E}_\gamma=\Rl^{2\dim(G)|E(\gamma)|}$. ${\cal E}_\gamma$ is equipped
with the norm 
\be \label{3.15a}
||\{x_e,y_e\}_{e\in E(\gamma)}||_{\rho,\sigma}:=
\sqrt{\sum_{e\in E(\gamma)} [\rho_e^{ij} x^i_e x^j_e+\sigma^e_{ij} y^i_e 
y^j_e]} 
\ee
where $\rho_e^{ij},\sigma^e_{ij}$ are metrics of Euclidean signature for each
$e$. Obviously, then $M_\gamma$ is a proper subset of $\bar{M}_\gamma$.\\
Remark :\\
A connection between (\ref{3.15a}) and (\ref{3.2}) can be given 
on certain graphs as follows : The idea is that (\ref{3.15a}) is a 
discretization of 
(\ref{3.2}) so that they eqaul each other in the limit of an infinitely
fine graph. Consider for simplicity $\Sigma=\Rl^3$ (the generalization 
to arbitrary $\Sigma$ is straightforward with the tools of the next 
subsection) and consider a lattice $\gamma$ of regular cubic topology. Then 
edges can be labelled as $t\mapsto e_I(v,t)$ where $v=e_I(v,0)$ is a vertex  
and $I=1,..,D$. Let
edges be images of the interval $t\in[0,\epsilon]$ and define $Y^a_I(v)=
\dot{e}^a_I(v,0),\;n_a^I(v)=\frac{1}{(D-1)!}
\epsilon_{a b_1..b_{D-1}} \epsilon^{I J_1..J_{D-1}}
Y^{b_1}_{J_1}(v)..Y^{b_{D-1}}_{J_{D-1}}(v)$. We now choose 
$\rho^e_{ij}=\delta_{ij}f(e),\;\sigma_e^{ij}=\delta^{ij}g(e)$
for some functions $f,g:\;E(\gamma)\mapsto \Rl$ and choose the $e_I(v)$ in 
such a way that 
$$
\sum_I f(e_I(v))) Y^a_I(v) 
Y^b_I(v)=\epsilon^{D-2}(\sqrt{\det(\rho)}\rho^{ab})(v) \mbox{ and }
\sum_I g(e_I(v)) n_a^I(v) 
n_b^I(v)=\epsilon^{-(D-2)}(\sigma_{ab}/\sqrt{\det(\sigma)})(v) 
$$
the metrics of (\ref{3.2}) result. It is then easy to see that with 
$x^e_i:=-\frac{2}{N}\mbox{tr}(\tau_i h_e(A) h_e(A')^{-1}),\;
y_e^i:=P^e_i(A,E)-P^e_i(A',E')$ the resulting metric (\ref{3.15a}) 
converges to the metric (\ref{3.2}).\\
\\
In order to proceed and to give $M_\gamma$ a symplectic structure 
derived from that of $M$ one must regularize the elementary 
functions so that one can use the symplectic structure $\Omega$, then 
study the limit in which the regulator is removed and hope that the result
is a well-defined symplectic structure $\Omega_\gamma$. We choose the 
following regularization :\\
Given an edge $e$ we define a tube $T^e$ around $e$ to be a foliation 
of $D-1$dimensional surfaces which are topologically discs. Let
$f_\epsilon :\; D \mapsto T^e_\epsilon$ be a one-parameter
family of smooth functions which tends to the $\delta$-distribution
on the unit disc $D$. Recall that the holonomy of a smooth connection can 
be written as the path ordered exponential
\be \label{3.16}
h_e(A)=1+\sum_{n=1}^\infty\int_0^1 dt_n\int_0^{t_n} dt_{n-1}..
\int_0^{t_2}dt_1 A(t_1)..A(t_n)
\ee
where $A(t)=\dot{e}^a(t) A_a^i(e(t))\tau_i/2$. We define the holonomy
smeared over the tube $T^e_\epsilon$ by
\be \label{3.17}
h^\epsilon_e(A)=1+\sum_{n=1}^\infty\int_0^1 dt_n\int_0^{t_n} dt_{n-1}..
\int_0^{t_2}dt_1 \int_D d^{D-1}y_n f_\epsilon(y_n)..\int_D 
d^{D-1}y_1 f_\epsilon(y_1) A_{y_1}(t_1)..A_{y_n}(t_n) 
\ee
where $A_y(t)=\dot{e}^a_y(t)A_a^i(e_y(t))\tau_i/2$ and 
$D\times [0,1]\mapsto T^e;\;(y,t)\mapsto e_y(t)$ defines the tube $T^e$
with the convention that $e_0(t)=e(t)$.

Likewise, let $R^e$ be a region foliated by surfaces diffeomorphic to 
$S^e$. Let $X:\; V\mapsto S^e;\;u:=(u_1,..,u_{D-1}))\mapsto 
X(u)$
be a parameterization of $S^e$ where $V$ is an open submanifold of 
$\Rl^D$ of dimension $D-1$. Furhermore, let 
$[-1,1]\times V\mapsto R^e;\; (s,u)\mapsto X_s(u)$ define $R^e$ with the 
convention that $X_0(u)=X(u)$. Let $g_\epsilon$ be a one parameter family
of smooth functions which tends to the $\delta$-distribution on the interval
$[-1,1]$ as $\epsilon\to 0$. Also, define $\rho_e^s(x)$ to be paths 
in $X_s(V)=S^s_e$ between $p_e(s)=X_s(0)=e\cap X_s(V)$ and $x\in X_s(V)$ 
and let $e_s$ be a 
reparameterization of $e$ such that $e_s(1/2)=p_e^s$. Here we assume w.l.g.
that all the surfaces $S^s_e,s\in[-1,1]$ satisfy the conditions that the 
surface $S_e$ has to satisfy. We can now define
\be \label{3.18}
P^e_{i,s}(A,E):=-\frac{1}{N}
\mbox{tr}(\tau_i h_{e_s}(0,1/2)[\int_{S^s_e} h_{\rho_{e_s}(x)} \ast E(x) 
h_{\rho_{e_s}(x)}^{-1}] h_{e_s}(0,1/2)^{-1})
\ee
and then 
\be \label{3.19}
P^{e\epsilon}_i:=\int_{-1}^1 ds g_\epsilon(s) P^e_{i,s}
\ee
Notice that the holonomies involved in (\ref{3.19}) remain unsmeared 
as compared to (\ref{3.17}). We could improve this by an additional smearing,
preserving gauge covariance,
but it would just blow up the subsequent calculations and would not 
change the end result. The careful reader is invited to fill in the 
missing details. 

Apart from these details, the functions (\ref{3.17}) and (\ref{3.19})
are now written as functions of the variables $F(A),E(f)$ where
$F,f$ are of the form $F^a_i(x)=\chi_{T^e}(x)(f_\epsilon(y)\dot{e}^a_y(t)
\delta_i^j)_{x=e_y(t)}$ and 
$f_a^i(x)=\chi_{R^e}(x)(g_\epsilon(s)\epsilon_{ab_1..b_{D-1}}
X^{b1}_{s,u_1}(u)..X^{b2}_{s,u_{D-1}}(u)\delta_i^j)_{x=X_s(u)}$ 
for some $j$ and are thus certainly elements of $\cal S$. It follows that 
the smeared functions are functionally differentiable.
Moreover, by construction, the smeared objects converge pointwise on $M$
to the unsmeared objects, that is  
\be \label{3.20}
\lim_{\epsilon\to 0} |(h_e^\epsilon(A))_{AB}-(h_e(A))_{AB})|=
\lim_{\epsilon\to 0} |P^{e\epsilon}_i(A,E)-P^e_i(A,E)|=0
\ee
for all $(A,E)\in M,\;i=1,..,\dim(G),\; A,B$ where $A,B,..$ denote group 
indices. 
\begin{Theorem} \label{th3.2}
The smeared variables allow us to define the following bracket 
$\{.,.\}_\gamma$ on $M_\gamma$ :
\ba \label{3.21a}
\{h_e,h_{e'}\}_\gamma &:=&
\lim_{\epsilon_1,\epsilon_2\to 
0}\{h^{\epsilon_1}_e,h^{\epsilon_2}_{e'}\}=0\\
\label{3.21b}
\{P^e_i,h_{e'}\}_\gamma &:=&
\lim_{\epsilon_1\to 0}\lim_{\epsilon_2\to 0} 
\{P^{e\epsilon_1}_i,h^{\epsilon_2}_{e'}\}=
\delta^e_{e'} \frac{\tau_i}{2}h_e\\
\label{3.21c}
\{P^e_i,P^{e'}_j\}_\gamma &:=&
\lim_{\epsilon_1,\epsilon_2\to 0}\{P^{e\epsilon_1}_i,P^{e'\epsilon_2}_j\}=
-\delta^{ee'}f_{ij}\;^k P^e_k
\ea
where $\{.,.\}$ is the bracket on $M$ and convergence is meant here and 
in what follows to be pointwise on $M$.
\end{Theorem}
Notice that we do not yet call $\{.,.\}_\gamma$ a Poisson bracket since 
we must check that it qualifies as a (strong) symplectic structure. This 
we will do in a separate step.\\
Proof of Theorem \ref{th3.2} :\\
{[}1.]\\
Recalling (\ref{3.13}) the first identity (\ref{3.21a}) follows easily 
from the fact that $\{h^{\epsilon_1}_e,h^{\epsilon_2}_{e'}\}=0$ at every 
finite $\epsilon_1,\epsilon_2$. It is for this reason that we do not have to 
smear the 
$h_{\rho_e^s(x)},h_{e_s}$ involved in (\ref{3.18}) in addition to $E$ in 
order 
to define the brackets, there would be no extra contribution due to this 
fact.\\
{[}2.]\\
The second identity (\ref{3.21b}) is significantly more involved. We 
first prove the following lemma.
\begin{Lemma} \label{la3.1}
For any $f_a^i\in {\cal S}$ and any path $e$ we have 
\be \label{3.22}
\{E(f),h_e\}:=\lim_{\epsilon\to 0} \{E(f),h^\epsilon_e\}
=\int_0^1 dt \dot{e}^a(t) f_a^i(e(t))h_e(0,t)\frac{\tau_i}{2}h_e(t,1)
\ee
\end{Lemma}
Proof of Lemma \ref{la3.1} :\\
We have by definition
\ba \label{3.23}
\{E(f),h^\epsilon_e(A)\}&=&
\sum_{n=1}^\infty\int_0^1 dt_n\int_0^{t_n} dt_{n-1}..
\int_0^{t_2}dt_1 \int_D d^{D-1}y_n f_\epsilon(y_n)..\int_D 
d^{D-1}y_1 f_\epsilon(y_1) \times\nonumber\\
&& \times \sum_{k=1}^n 
A_{y_1}(t_1)..\{E(f),A_{y_k}(t_k)\}.. A_{y_n}(t_n) \nonumber\\
&=& 
\sum_{n=1}^\infty\int_0^1 dt_n\int_0^{t_n} dt_{n-1}..
\int_0^{t_2}dt_1 \int_D d^{D-1}y_n f_\epsilon(y_n)..\int_D 
d^{D-1}y_1 f_\epsilon(y_1) \times\nonumber\\
&&\times \sum_{k=1}^n f_a^i(e_{y_k}(t_k))
\dot{e}^a_{y_k}(t_k)
A_{y_1}(t_1)..\frac{\tau_i}{2}.. A_{y_n}(t_n) 
\ea
Relabel $T_1=t_1,..,T_{k-1}=t_{k-1},t=t_k,T_{k}=t_{k+1},..,T_{n-1}=t_n$
and $z_1=y_1,..,z_{k-1}=y_{k-1},y=y_k,z_{k}=y_{k+1},..,z_{n-1}=y_n$
then (\ref{3.23}) becomes 
\ba \label{3.24}
\{E(f),h^\epsilon_e(A)\}&=&
\sum_{n=1}^\infty\sum_{k=1}^n \int_0^1 dT_{n-1}\int_0^{T_{n-1}} dT_{n-2}..
\int_0^{T_{k+1}} dT_k\int_0^{T_k} dt\int_0^t dT_{k-1}..\int_0^{T_2} dT_1
\times\nonumber\\
&& \times
\int_D d^{D-1}y f_\epsilon(y) f_a^i(e_y(t))\dot{e}^a_y(t)
\int_D d^{D-1}z_{n-1} f_\epsilon(z_{n-1})..\int_D 
d^{D-1}z_1 f_\epsilon(z_1) 
\times\nonumber\\ && \times
\sum_{k=1}^n 
A_{z_1}(T_1)..\frac{\tau_i}{2}.. A_{z_{n-1}}(T_{n-1}) 
\nonumber\\
&=&
\sum_{n=1}^\infty\sum_{k=1}^n 
\int_0^1 dt \int_D d^{D-1}y f_\epsilon(y) f_a^i(e_y(t))\dot{e}^a_y(t)
\times\nonumber\\
&&\times
\int_t^1 dT_{n-1}\int_t^{T_{n-1}} dT_{n-2}..
\int_t^{T_{k+1}} dT_k\int_0^t dT_{k-1}\int_0^{T_{k-1}}dT_{k-2}..\int_0^{T_2} 
dT_1 
\times\nonumber\\
&& \times \int_D d^{D-1}z_{n-1} f_\epsilon(z_{n-1})..\int_D 
d^{D-1}z_1 f_\epsilon(z_1) \sum_{k=1}^n
A_{z_1}(T_1)..\frac{\tau_i}{2}.. A_{z_{n-1}}(T_{n-1}) 
\ea
where in the last step we have used the fact that 
$1\ge T_{n-1}\ge..\ge T_k\ge t\ge T_{k-1}\ge ..\ge T_1\ge 0$
in order to make the range of integration of $t$ independent of the $T_k$.
We can now easily take the limit $\epsilon\to 0$ with the result
\ba \label{3.25}
&& \sum_{n=1}^\infty\sum_{k=1}^n 
\int_0^1 dt f_a^i(e(t))\dot{e}^a(t)
\int_t^1 dT_{n-1}\int_t^{T_{n-1}} dT_{n-2}..
\nonumber\\
&&
..\int_t^{T_{k+1}} dT_k\int_0^t dT_{k-1}\int_0^{T_{k-1}}dT_{k-2}..\int_0^{T_2} 
\sum_{k=1}^nA_{z_1}(T_1)..\frac{\tau_i}{2}.. A_{z_{n-1}}(T_{n-1}) 
\ea
and writing out the path product identity for holonomies
$h_e(0,t)h_e(t,1)=h_e(0,1)=h_e$ in the path ordered form (\ref{3.16})
this collapses indeed to (\ref{3.22}).\\
$\Box$\\
Let us now prove the second identity (\ref{3.21b}). Let us write 
$P^{e\epsilon}_i$ in the form $E(f)$ by choosing
\ba \label{3.26}
(f^\epsilon_i)_a^j(x) &=&-\int_{-1}^1 ds \int_V d^{D-1}u \delta(x,X_s(u))
\frac{1}{N} g_\epsilon(s)
\epsilon_{ab_1..b_{D-1}}X^{b_1}_{s,u_1}(u) ..X^{b_{D-1}}_{s,u_{D-1}}(u) 
\times\nonumber\\ &&\times
\mbox{tr}(\tau_i h_{e_s}(0,1/2)h_{\rho_{e_s}(X_s(u))} 
\tau_j h_{\rho_{e_s}(X_s(u))}^{-1} h_{e_s}(0,1/2)^{-1})
\ea
From [1.] it is clear that, although $f_a^j$ depends on $A$, as far as 
(\ref{3.21b}) is concerned, we can treat it as 
a numerical function. By Lemma \ref{la3.1} we then have
\be \label{3.27}
\{P^e_i,h_{e'}\}_\gamma=\lim_{\epsilon\to 0} \{E(f^\epsilon_i),h_{e'}\}
=\lim_{\epsilon\to 0} \int_0^1 dt \dot{e}^{a\prime}(t) 
(f^\epsilon_i)_a^j(e'(t)) h_{e'}(0,t)\frac{\tau_j}{2}h_{e'}(t,1) 
\ee
Suppose first that $e\not=e'$. Then, no matter how complicated 
$\gamma,P_\gamma$ look, for sufficiently small $\epsilon$ the edge
$e'$ does not intersect the region $R^e_\epsilon$ and thus (\ref{3.27}) 
vanishes.
If $e=e'$ then $e$ intersects $R^e_\epsilon$ for any
value of $\epsilon$ and we find
\ba \label{3.28} 
&& \{P^e_i,h_e\}_\gamma=-\lim_{\epsilon\to 0}
\frac{1}{N} \int_{-1}^1 ds g_\epsilon(s) \int_0^1 dt \int_V d^{D-1}u 
\delta(e(t),X_s(u))\dot{e}^a(t)\times \nonumber\\
&& \times 
\epsilon_{ab_1..b_{D-1}}X^{b_1}_{s,u_1}(u) ..X^{b_{D-1}}_{s,u_{D-1}}(u) 
\mbox{tr}(\tau_i h_{e_s}(0,1/2)h_{\rho_{e_s}(X_s(u))} 
\times\nonumber\\ && \times
\tau_j h_{\rho_{e_s}(X_s(u))}^{-1} h_{e_s}(0,1/2)^{-1})
h_e(0,t)\frac{\tau_j}{2}h_e(t,1) 
\ea
At fixed $s$ the only contribution of the integral over 
$(t,u)\in [0,1]\times V$ comes from the value $(t=t_s,u=0)$ 
since $S^s_e$ and $e$ intersect in the only point $p^s_e$
which is an interior point of $[0,1]\times V$ for sufficiently small 
$\epsilon$. Here $t_s$ is defined by $e_s(1/2)=e(t_s)$.
By definition of the orientation of the $X_s(V)$ we know that  
$\dot{e}^a(t) 
\epsilon_{ab_1..b_{D-1}}X^{b_1}_{s,u_1}(u) ..X^{b_{D-1}}_{s,u_{D-1}}(u)>0$
at $(t=t_s,u=0)$. Since $\rho_{e_s}(p_e^s)=p_e^s$, (\ref{3.28}) collapses to
\be \label{3.29} 
\{P^e_i,h_{e'}\}_\gamma=-\lim_{\epsilon\to 0}
\frac{1}{N} \int_{-1}^1 ds g_\epsilon(s) 
\mbox{tr}(\tau_i h_{e_s}(0,1/2)\tau_j h_{e_s}(0,1/2)^{-1})
h_{e_s}(0,1/2)\frac{\tau_j}{2}h_{e_s}(1/2,1) 
\ee
Since the integrand depends continuously on $s$ for any smooth connection
we see that pointwise
\be \label{3.30} 
\{P^e_i,h_{e'}\}_\gamma=-\frac{1}{N} 
\mbox{tr}(\tau_i h_e(0,1/2)\tau_j h_e(0,1/2)^{-1})
h_e(0,1/2)\frac{\tau_j}{2}h_e(1/2,1) 
\ee
Now consider the matrix $C_i=h_e(0,1/2)^{-1}\tau_i h_e(0,1/2)$ which is an 
element of the Lie algebra of $G$ because it is simply the transform of 
$\tau_i$ under the action of $h_e(0,1/2)^{-1}$ in the adjoint 
representation of $G$ on $Lie(G)$. Thus we can expand $C_i$ in the basis 
$\tau_j$, resulting in $C_i=-\mbox{tr}(C_i\tau_j)\tau_j/N$. Inserting 
this identity into (\ref{3.30}) gives the result claimed.\\
{[}3.]\\
Let us now turn to the final third identity (\ref{3.21c}). It is clear 
that for $e\not=e'$ and  
$\epsilon_1,\epsilon_2$ sufficiently small the regions 
$R_e^{\epsilon_1},R_{e'}^{\epsilon_2}$ are disjoint in which case the 
brackets vanish. Thus we only need to be concerned with the case $e=e'$.
Let us again use the convention (\ref{3.26}) and let us introduce the 
notation that for a function $f$ depending on $(A,E)$, the function
$\tilde{f}$ is numerically equal to $f$ but $\tilde{f}$ is considered to 
be independent of $(A,E)$. In other words, $\tilde{f}$ drops out of Poisson 
brackets on $M$ but $f$ does not necessarily. Then we have by the Leibniz
rule for $\Omega$
\ba \label{3.31}
&& \{P^{e\epsilon_1}_i,P^{e\epsilon_2}_j\}
=\{E(f^{\epsilon_1}_i),E(f^{\epsilon_2}_j)\}
\nonumber\\
&=&\{E(\tilde{f}^{\epsilon_1}_i),E(\tilde{f}^{\epsilon_2}_j)\}
+\{E(\tilde{f}^{\epsilon_1}_i),\tilde{E}(f^{\epsilon_2}_j)\}
+\{\tilde{E}(f^{\epsilon_1}_i),E(\tilde{f}^{\epsilon_2}_j)\}
+\{\tilde{E}(f^{\epsilon_1}_i),\tilde{E}(f^{\epsilon_2}_j)\}
\ea
The first term drops out by definition of $\Omega$ (recall (\ref{3.13}))
and the fourth by [1.] so that only the second and third term 
survive. More explicitly
\be \label{3.32}
\{P^{e\epsilon_1}_i,P^{e\epsilon_2}_j\}
=\int_\Sigma d^3x E^a_k(x)
[\{E(\tilde{f}^{\epsilon_1}_i),(f^{\epsilon_2}_j)_a^k(x))\}
-\{E(\tilde{f}^{\epsilon_2}_j),(f^{\epsilon_1}_i)_a^k(x))\}]
\ee
In order to compute (\ref{3.32}) in the limit $\epsilon_1,\epsilon_2\to 0$
we have to consider two types of terms, namely
$\{E(\tilde{f}^{\epsilon}_i),h_{\rho_{e_s}(X_s(u))}\}$ and 
$\{E(\tilde{f}^{\epsilon}_i),h_{e_s}(0,1/2)\}$
The first term is given, according to Lemma \ref{la3.1}, by 
\ba \label{3.33}
&& \int_0^1 dt \dot{\rho}^a_{e_s}(X_s(u),t) 
(\tilde{f}^\epsilon_i)_a^j(\rho_{e_s}(X_s(u))
h_{\rho_{e_s}(X_s(u))}(0,t)\frac{\tau_j}{2}h_{\rho_{e_s}(X_s(u))}(t,1)
\nonumber\\
& = & -\frac{1}{N}\int_0^1 dt 
h_{\rho_{e_s}(X_s(u))}(0,t)\frac{\tau_j}{2}h_{\rho_{e_s}(X_s(u))}(t,1)
\int_{-1}^1 dr g_\epsilon(r) \int_V d^{D-1}v 
\times\nonumber\\ &&\times
\delta(\rho_{e_s}(X_s(u),t),X_r(v))
\dot{\rho}^a_{e_s}(X_s(u),t) \epsilon_{ab_1..b_{D-1}}
X^{b_1}_{r,v_1}(v)..X^{b_{D-1}}_{r,v_{D-1}}(v)
\times\nonumber\\ && \times
\mbox{tr}(\tau_i h_{e_r}(0,1/2)h_{\rho_{e_r}(X_r(v))} 
\tau_j h_{\rho_{e_r}(X_r(v))}^{-1} h_{e_r}(0,1/2)^{-1})
\ea
Using the fact that $r\mapsto X_r(V)$ defines a foliation $R_e$ of surfaces
diffeomorphic to $S_e$ we see that the integral over $r,v$ is supported 
at the interior point $r=(s,v(s,t,u))$ of $[-1,1]\times V$ where 
$X_s(v(s,t,u))=\rho_{e_s}(X_s(u),t)$. The integral can be performed
with the result 
\ba \label{3.34}
&& -\frac{1}{N}\int_0^1 dt\; 
h_{\rho_{e_s}(X_s(u))}(0,t)\frac{\tau_j}{2}h_{\rho_{e_s}(X_s(u))}(t,1)
g_\epsilon(s) \times \nonumber\\ 
&& \times
[\frac{\dot{\rho}^a_{e_s}(X_s(u),t) \epsilon_{ab_1..b_{D-1}}
X^{b_1}_{r,v_1}(v)..X^{b_{D-1}}_{r,v_{D-1}}(v)}
{|\epsilon_{ab_1..b_{D-1}}
X^a_{r,r} X^{b_1}_{r,v_1}(v)..X^{b_{D-1}}_{r,v_{D-1}}(v)|}
\times\nonumber\\
&& \times
\mbox{tr}(\tau_i h_{e_r}(0,1/2)h_{\rho_{e_r}(X_r(v))} 
\tau_j h_{\rho_{e_r}(X_r(v))}^{-1} h_{e_r}(0,1/2)^{-1})]_{r=s,v=v(s,t,u)}
\ea
Notice that the denominator in (\ref{3.34}) is bounded away from zero
as the curve $s\mapsto X_s(v)$ for any fixed $v$ is transversal to $X_s(V)$. 
Now 
$\rho_{e_s}(X_s(u),t)=(X_r(v))_{r=s,v=v(s,t,u)}$ 
for any $t\in[0,1]$, thus 
$\dot{\rho}_{e_s}(X_s(u),t)=\sum_{k=1}^{D-1}(X_{r,v_k}(v))_{r=s,v=v(s,t,u)}
\frac{dv_k(s,t,u)}{dt}$, thus the integrand of (\ref{3.34}) vanishes for any
finite $\epsilon$. 

Thus there will be no contribution from the holonomies along the 
$\rho_{e_s}(X_s(u))$ and we can focus on the second term 
$\{E(\tilde{f}^{\epsilon}_i),h_{e_s}(0,1/2)\}$ mentioned.
Again, according to Lemma \ref{la3.1}, it is given by
\ba \label{3.35}
&& \int_0^{1/2} dt \dot{e}^a_s(t)
(\tilde{f}^\epsilon_i)_a^j(e_s(t))
h_{e_s}(0,t)\frac{\tau_j}{2}h_{e_s}(t,1/2)
\nonumber\\
& = & -\frac{1}{N}\int_0^{1/2} dt \;
h_{e_s}(0,t)\frac{\tau_j}{2}h_{e_s}(t,1/2)
\int_{-1}^1 dr g_\epsilon(r) \int_V d^{D-1}v 
\delta(e_s(t),X_r(v))
\dot{e}^a_s(t)\times\nonumber\\
&& \times\epsilon_{ab_1..b_{D-1}}
X^{b_1}_{r,v_1}(v)..X^{b_{D-1}}_{r,v_{D-1}}(v)
\mbox{tr}(\tau_i h_{e_r}(0,1/2)h_{\rho_{e_r}(X_r(v))} 
\tau_j h_{\rho_{e_r}(X_r(v))}^{-1} h_{e_r}(0,1/2)^{-1})
\ea
This time we will perform the $t,v$ integral to cancel the 
$\delta$-distribution. Thus we may as well perform the limits 
$\epsilon_1,\epsilon_2\to 0$ and cancel the $s$ and $r$ integrals first 
with the result that we are left with
\ba \label{3.36}
&& -\frac{1}{N}\int_0^{1/2} dt 
h_e(0,t)\frac{\tau_j}{2}h_e(t,1/2)
\int_V d^{D-1}v 
\delta(e(t),X(v))
\dot{e}^a(t)
\times\nonumber\\
&& \times \epsilon_{ab_1..b_{D-1}}
X^{b_1}_{,v_1}(v)..X^{b_{D-1}}_{,v_{D-1}}(v)
\mbox{tr}(\tau_i h_e(0,1/2)h_{\rho_e(X(v))} 
\tau_j h_{\rho_e(X(v))}^{-1} h_e(0,1/2)^{-1})
\ea
The integrand is supported at $t=1/2, v=0$ which is now a {\it boundary}
point of the interval $[0,1/2]$ which is why the $\delta$ distribution
picks up a factor of $1/2$ as compared to the analogous 
argumentation in [2.]. This leads to the result
\be \label{3.37}
-\frac{1}{2N}
h_e(0,1/2)\frac{\tau_j}{2}\mbox{tr}(\tau_i h_e(0,1/2)\tau_j  h_e(0,1/2)^{-1})
\nonumber\\
=\frac{1}{4}\tau_i h_e(0,1/2)
\ee
Putting things together and using $\delta g^{-1}=-g^{-1}\delta g g^{-1}$
we find 
\ba \label{3.38}
\{P^e_i,P^e_j\}_\gamma
&=& \int_\Sigma d^3x E^a_k(x)
[-\frac{1}{N} \int_V d^{D-1}u \delta(x,X(u))
\epsilon_{ab_1..b_{D-1}}X^{b_1}_{u_1}(u) ..X^{b_{D-1}}_{u_{D-1}}(u)]
\times\nonumber\\
&\times& [\mbox{tr}(\tau_j 
\{\frac{1}{4}\tau_i h_e(0,1/2)\}
h_{\rho_e(X(u))} 
\tau_k h_{\rho_e(X(u))}^{-1} h_e(0,1/2)^{-1})\nonumber\\
&& -
\mbox{tr}(\tau_j h_e(0,1/2)
h_{\rho_e(X(u))} 
\tau_k h_{\rho_e(X(u))}^{-1} 
h_e(0,1/2)^{-1}
\{\frac{1}{4}\tau_i h_e(0,1/2)\}
h_e(0,1/2)^{-1})
\nonumber\\
&& - i\leftrightarrow j]\nonumber\\
&=& -\frac{1}{4N}\int_{S_e}  (\ast E)_k(x)\times\nonumber\\
&\times& [\mbox{tr}(\tau_j\tau_i h_e(0,1/2)
h_{\rho_e(x)} 
\tau_k h_{\rho_e(x)}^{-1} h_e(0,1/2)^{-1})\nonumber\\
&& -
\mbox{tr}(\tau_j h_e(0,1/2)h_{\rho_e(x)} 
\tau_k h_{\rho_e(x)}^{-1} h_e(0,1/2)^{-1}
\tau_i) \nonumber\\
&& - i\leftrightarrow j\nonumber]\\
&=& -\frac{1}{4N}\int_{S_e}  
[2\mbox{tr}([\tau_j,\tau_i] h_e(0,1/2)
h_{\rho_e(x)} 
\ast E(x)h_{\rho_e(x)}^{-1} h_e(0,1/2)^{-1})]\nonumber\\
&=& \frac{f_{ij}\;^k}{N}\int_{S_e}  
\mbox{tr}(\tau_k h_e(0,1/2)
h_{\rho_e(x)} 
\ast E(x)h_{\rho_e(x)}^{-1} h_e(0,1/2)^{-1})\nonumber\\
&=& -f_{ij}\;^k P^e_k
\ea
as claimed.\\
$\Box$\\
Notice that the factor of $1/2$ that appeared in (\ref{3.38}) is also 
required if the bracket $\{.,.\}_\gamma$ is to satisfy the Leibniz rule :
$\{P^e_\gamma,h_e\}_\gamma=
\{P^e_\gamma,h_e(0,1/2)\}h_e(1/2,1)
+h_e(0,1/2)\{P^e_\gamma,h_e(1/2,1)\}$ which is consistent 
with (\ref{3.21b}) if indeed
$\{P^e_\gamma,h_e(0,1/2)\}=\tau_i h_e(0,1/2)/4,
\{P^e_\gamma,h_e(1/2,1)\}=h_e(0,1/2)^{-1}\tau_i h_e/4$.\\
\\
\begin{Theorem} \label{th3.3}
The bracket $\{.,.\}_\gamma$ satisfies the Jacobi identity, moreover, it 
defines a non-degenerate two-form on $M_\gamma$, that is, it is a symplectic 
structure.
\end{Theorem}
Proof of Theorem \ref{th3.3} :\\
i) {\it Jacobi identity} :\\
There are four kinds of double-brackets to check corresponding to 
$n$ momenta and $3-n$ holonomies appearing with $n=0,1,2,3$.\\
$n=0)$ :\\
This case is trivial  
\be \label{3.38a}
\{(h_e)_{AB},\{(h_{e'})_{CD},(h_{e^{\prime\prime}})_{EF}\}_\gamma\}_\gamma
+\mbox{cyclic}=0
\ee
since already the inner bracket vanishes by (\ref{3.21a}).\\
$n=1)$ :\\
Also this case is trivial
\be \label{3.39}
\{(h_e)_{AB},\{(h_{e'})_{CD},P^{e^{\prime\prime}}_i\}_\gamma\}_\gamma+
\mbox{cyclic}=0
\ee
because either the inner bracket already vanishes or the inner 
bracket gives a function depending only on holonomies by (\ref{3.21b}) 
and so the outer bracket is of the type (\ref{3.21a}) and vanishes.\\
$n=2)$ :\\
This is the first non-trivial case and it is quite remarkable that
the signs and numerical factors in (\ref{3.21a})-(\ref{3.21c}) come  
in precisely the right way out of the regularized derivation of theorem 
\ref{th3.2} :
\ba \label{3.40}
&& \{h_e,\{P^{e'}_i,P^{e^{\prime\prime}}_j\}_\gamma\}_\gamma
+\{P^{e'}_i,\{P^{e^{\prime\prime}}_j,h_e\}_\gamma\}_\gamma
+\{P^{e^{\prime\prime}}_j,\{h_e,P^{e'}_i\}_\gamma\}_\gamma
\nonumber\\
&=& 
-\delta^{e'e^{\prime\prime}}f_{ij}\;^k\{h_e,P^{e'}_k\}_\gamma
+\delta^{e^{\prime\prime}}_e \{P^{e'}_i,\frac{\tau_j}{2}h_e\}_\gamma
-\delta^{e'}_e\{P^{e^{\prime\prime}}_j,\frac{\tau_i}{2}h_e\}_\gamma
\nonumber\\
&=& \delta^{e'e^{\prime\prime}} \delta^{e'}_e
(f_{ij}\;^k \frac{\tau_k}{2}h_e
+\frac{\tau_j\tau_i}{4}h_e
-\frac{\tau_i\tau_j}{4}h_e)
\nonumber\\
&=& \frac{1}{4}\delta^{e'e^{\prime\prime}} \delta^{e' e}
(2f_{ij}\;^k \tau_k+ [\tau_j,\tau_i])h_e =0
\ea
by definition of the structure constants.\\
$n=3$ :\\
This case is again easy, it just relies on the Jacobi identity for the 
generators of the Lie algebra of $G$ or, equivalently, for its structure 
constants :
\be \label{3.41}
\{P^e_i,\{P^{e'}_j,P^{e^{\prime\prime}}_k\}_\gamma\}_\gamma+\mbox{cyclic}
=\delta^{e'e^{\prime\prime}} \delta^{e'}_e
(f_{jk}\;^l f_{il}\;^m+\mbox{cyclic})P^e_m=0
\ee
ii) {\it Non-degeneracy} \\
Obviously, by (\ref{3.21a})-(\ref{3.21c}) the symplectic structure 
$\Omega_\gamma$, 
if it exists, is diagonal with respect to the edge label,
\be \label{3.42}
\Omega_\gamma=\sum_{e\in E(\gamma)} \Omega_e
\ee
where each $\Omega_e$ is isomorphic with a Poisson structure $\Omega_G$ on 
the cotangent bundle $T^\ast G$ given by $\{h_{AB},h_{CD}\}_G=0,\;
\{P_i,h\}_G=\tau_i h/2,\; \{P_i,P_j\}_G=-f_{ij}\;^k P_k$. Thus in order 
to show regularity of $\Omega_\gamma$ it will be necessary to show
regularity of $\Omega_G$, that is, $\Omega_G$ is not only a Poisson 
structure but actually a symplectic structure.\\
To see that $\Omega_G$ is everywhere nondegenerate on $G$ consider the atlas
$(U_\alpha,\phi_\alpha)_{\alpha\in I}$ of $G$ given by open neighbourhoods 
$U_\alpha$ containing some point $h_\alpha\in G$ and charts defined by
$\phi_\alpha^{-1}:=\exp:\;V_\alpha\subset \Rl^{\dim(G)}\mapsto 
U_\alpha;\;(\theta^j_\alpha)\mapsto \exp(\theta^j_\alpha \tau_j/2)h_\alpha
$. Obviously, the $\theta_\alpha^j$ serve as local coordinates of $G$.
Let now $h\in G$ be given, then there exists $\alpha\in I,
(\theta_\alpha^j)\in V_\alpha$ such that 
$h=\exp(\theta_\alpha^j\tau_j/2)h_\alpha$. By choosing the size of the 
index set $I$ high enough we can assume that the range of 
each $\theta_\alpha^j$ is contained in an open interval containing zero
as small as we wish. Let us now expand $h$ in powers of $\theta_\alpha^j$
in the relation $\{P_j,h\}_G=\tau_j h/2$ then we find by comparing powers 
that $\{P_j,\theta^k\}=\delta_j^k+O(\theta)$ where $O(\theta)$ is a 
bounded function vanishing linearly in $\theta$. We conclude that the bracket
when expressed in terms of the coordinates $P_j,\theta_j$ has locally
the form of a block matrix consisting of four blocks of 
$\dim(G)\times\dim(G)$ matrices with the off-diagonal blocks given
by plus/minus the identity matrix plus corrections of order $\theta_\alpha^j$
and the diagonal blocks consist of a zero matrix and of the matrix 
with $ij$-entries given by $f_{ij}\;^k P_k$. This matrix 
is obviously invertible on $U_\alpha$ for any $\alpha$, proving our claim.\\
Finally we must show that $\Omega_\gamma$ is a surjection, that is,
for any $(x_e,y_e)_{e\in E(\gamma)}\in{\cal E}'_\gamma$ there exists 
$(x'_e,y'_e)_{e\in E(\gamma)}\in{\cal E}_\gamma$ such that
$(x'_e,y'_e)\cdot \Omega_e=(x_e,y_e)$. Since ${\cal E}_\gamma$ is actually
a Hilbert-manifold, we find $(\tilde{x}_e,\tilde{y}_e)\in {\cal E}_\gamma$
such that $(x_e,y_e)=(\tilde{x}_e\cdot \rho_e,\tilde{y}_e\cdot \sigma_e)$
thus the solution of our problem is given by 
$(x'_e,y'_e)=(\tilde{x}_e\cdot \rho_e,\tilde{y}_e\cdot \sigma_e)\cdot
(\Omega_e)^{-1}$ ($\Omega_e$ interpreted as a matrix written pointwise in 
$M_\gamma$) 
provided we can show that this defines an element of ${\cal E}_\gamma$.
This is somewhat non-trivial because $\Omega_e$ depends on $P^e_i$ 
which, a priori, can take arbitrarily large values. However, the 
normalizability of 
this vector follows from (\ref{3.15a}) which implies that in fact $P^e_i$ 
must be uniformly bounded in $e$. \\
$\Box$\\ 
In the appendix we show that $\Omega_G$ and therefore $\Omega_\gamma$
is even exact in the case of $G=U(1), SU(2)$ which one can probably prove 
also for general $G$ and we leave this as a future project.

\subsubsection{Continuum Phase Space from the Graph Phase Space}
\label{s3.2.2}

In the previous subsection we established how the symplectic manifold 
$(M_\gamma,\Omega_\gamma)$ can be derived from the symplectic manifold
$(M,\Omega)$ for every $\gamma\in\Gamma$. In this subsection we wish to
show the opposite : in the limit that $\gamma$ grows ad infinitum 
{\it in a prescribed way} we find 
that $(M,\Omega)$ can be derived from $(M_\gamma,\Omega_\gamma)$. This 
lies at the heart of later constructions in which we use 
$(M_\gamma,\Omega_\gamma)$ as our starting point for quantization. 
Namely, as the formulation of the theory in terms of $M_\gamma$ is a 
certain kind of discretization, the result just stated means that the 
continuum limit exists and is the expected one on the classical level.
On the other hand, the symplectic manifold $(M_\gamma,\Omega_\gamma)$
is straightforward to quantize on a Hilbert space 
$({\cal H}_\gamma,<.,.>_\gamma)$ and the classical limit of this  
quantization is easily shown to give back $(M_\gamma,\Omega_\gamma)$.
In other words, we can establish the chain of limits 
$({\cal H}_\gamma,<.,.>_\gamma)\to_{\hbar\to 0}(M_\gamma,\Omega_\gamma)
\to_{\gamma\to\infty} (M,\Omega)$ and the interesting question of the 
existence of the opposite limit will be subject of our companion paper
\cite{23f}.\\
\\
In order to show that we can recover the continuum theory from the discrete
one in the limit of infinite graphs we consider a certain one-parameter 
family of cubic lattices $\epsilon\mapsto\gamma_\epsilon$ where $\epsilon$
is associated with the length of the edges or links of the graph with 
respect
to a certain background metric and the limit $\epsilon\to 0$ corresponds to 
sending the graph to the continuum $\Sigma$. More precisely, we have 
the following :\\
The manifold $\Sigma$ is described by an atlas of charts
$(U_\iota,X_\iota)_{\iota\in{\cal I}}$ where $U_\iota$ is an open region in 
$\Sigma$ and $X_\iota^a\; :\; V_\iota\subset \Rl^D\mapsto U_\iota, 
(t^1,..,t^D)\mapsto X_\iota(t)$ is a
local trivialization of $U_\iota$, that is, a smooth orientation
preserving diffeomorphism. 
Consider an arbitrary but fixed decomposition $\cal R$ of $\Sigma$ into 
mutually disjoint, except for common boundary points, compact regions $R$ 
which is fine enough such that 
every $R$ lies in the domain of a chart and choose
$\iota(R)\in {\cal I}$ to be such that $R\in U_{\iota(R)}$
(at this point we do not even need the cover $\{ U_i\}$ to be locally 
finite or
$\Sigma$ to be paracompact although this is an assumption which goes into
the definition of $\Gamma^\omega_\sigma$). 
Without loss of generality we can assume 
that each $R$ is diffeomorphic to a polyhedron of $\Rl^D$ and we fix for 
each of its faces an orientation. Also, if $\Sigma$ is not compact we 
take a refinement of the atlas if necessary such that each $R$ has finite
Lebesgue measure.

Now $R$ is the image under $X_R:=X_{\iota(R)}$ of a compact region 
$V_R$, in fact a polyhedron, in 
$\Rl^D$ which can always be decomposed, for sufficiently small $\epsilon$,  
into regular cubes of volume $\epsilon^D$ with respect to the Euclidean 
metric of $\Rl^D$ possibly up to a subset near the the boundary of 
$X^{-1}_R(R)$. Let then $C_R$ be the union of those cubes which fit 
into $V_R$. 

Now each region $C_R$ is filled exactly with cubes of volume $\epsilon^D$
and these cubes define a regular, oriented cubic lattice $\gamma^0_R$ in 
$C_R$, the orientation of the edges is chosen to be such 
that each of them points in the positive coordinate axis direction of
$\Rl^D$. Let $P^0_{\gamma^0_R}$ be the dual 
decomposition
of $V_R$ obtained as follows : choose for each edge 
$e^0_R$ of $\gamma^0_R$ 
the open face $S^0_{e^0_R}$ to be the regular cubic hyperplane orthogonal to 
$e^0_R$ of area $\epsilon^{D-1}$
which cuts $e^0_R$ in the middle, is pierced by $e^0_R$ in its center
and carries the orientation defined by choosing the tangent
direction of $e^0_R$ to be the direction of its unit normal vector.
The collection of all these faces $S^0_{e^0_R}$ can be completed uniquely 
to the unique, minimal cubic polyhedronal complex $C'_R$ which contains  
all of them. We assume w.l.g. that $C'_R$ fits into
$V_R$ (decrease the size of $C_R$ by deleting some cubes 
on its boundary if necessary). Notice that $C'_R$ necessarily covers 
$C_R$. We complete the polyhedronal
decomposition of $V_R$ by choosing $X^{-1}_R(R)-C'_R$ as the final
polyhedron to cover $V_R$. 
Since the boundary of $V_R$ is already oriented this obviously 
defines an 
oriented polyhedronal decomposition of $X^{-1}_R(R)$ dual to 
$\gamma^0_R$.

Finally, we consider the images
$\gamma_R=X_{\iota(R)}(\gamma^0_R),\; S_e=X_{\iota(R)}(S^0_{e^0_R})
\; P_{\gamma_R}=X_{\iota(R)}(P^0_{\gamma^0_R})$
and the corresponding variables $h_e^R,P^{e,R}_i$ for each
$R\in {\cal R}$. The unions $\gamma_\epsilon:=\cup_{R\in {\cal R}}\; 
\gamma_R$ 
and $P_{\gamma_\epsilon}:=\cup_{R\in {\cal R}} P^0_{\gamma_R}$ define an 
oriented graph in $\Sigma$ and an oriented decomposition of $\Sigma$.
It is not yet a decomposition dual to $\gamma$ since it is not minimal :
it becomes minimal if we remove all the boundaries $\partial R$. Let the
resulting dual decomposition also be denoted by $P_{\gamma_\epsilon}$.

Notice that the region $\Sigma-[\cup_{R\in{\cal R}}X_R(C'_R)]$ does 
not contain any piece of $\gamma_\epsilon$, however, its Lebesgue measure 
vanishes in the limit $\epsilon\to 0$ because it tends to 
$\cup_R \partial R$.
We could avoid this by adding edges to $\gamma$ connecting the
$\gamma_R$ which are contained in neigbouring $R$ but the resulting lattice 
may not be of cubic topology any longer. Since we will not need those edges 
for the sake of our argument, we will leave things as they are. 

Let us fix a specific $R$ and define 
$v_R:=X_R(v^0_R),\;e_{RI}(v):=X_R(e^0_{RI}(v^0)),\; 
S^I_R(v):=X_R(S^0_{e^0_{RI}(v^0_R)})$ 
respectively to be the image under $X^R$ of a vertex, edge and 
face of $\gamma^0_R$ respectively. Here $e^0_{RI}(v^0_R)$ denotes 
the straight line into the positive $I-$direction between the 
vertices $v^0_R$ and $v^0_R+\epsilon b_I$ where $\{b_I\}_{I=1}^D$ denotes 
the standard oriented orthonormal basis of $\Rl^D$ (sometimes 
$v^0_R+\epsilon b_I$ is not a vertex of $\gamma^0_R$ in which case we set
$e^0_{RI}(v^0_R)=v^0_R=S^0_{e^0_{RI}(v^0_R)}$).
Consider also for any $x\in R$ the vector fields $Y^a_{RI}(x):=
X^a_{R,I}(t)_{x=X_R(t)}$ and co-vector densities of weight minus one
$n^I_{Ra}(x):=\frac{1}{(D-1)!}\epsilon_{ab_1..b_{D-1}}\epsilon^{IJ_1..J_{D-1}} 
Y^{b_1}_{RJ_1}(x)..Y^{b_{D-1}}_{RJ_{D-1}}(x)$.
Since $X_R:\;V_R\mapsto R$ is an orientation preserving diffeomorphism 
it is clear that $\det((Y))=n^I_a Y^a_I >0$ everywhere in $R$. 

We define now as in (\ref{3.14}) for every vertex $v$ of $\gamma_R$
the functions (we drop the label $R$)  
\ba \label{3.42a}
h_I(v) &:=& h_{e_I(v)}(A) \\
P^I_i(v) &:=& -\frac{1}{N}
\mbox{tr}(\tau_i h_{e_I(v)}(0,1/2)[\int_{S^I(v)} h_{\rho_{e_I(v)}(x)} 
\ast E(x) h_{\rho_{e_I(v)}(x)}^{-1}] h_{e_I(v)}(0,1/2)^{-1})(A,E)
\nonumber
\ea
which defines a map $D_\epsilon :\; M\mapsto M_\gamma$.

On the other hand, consider now the following functions on $M_\gamma$  
($v$ is again a vertex of $\gamma_R$ and we drop the label $R$) 
\ba \label{3.43}
A^{(\epsilon)i}_a(v) &:= & -2\frac{n_a^I}{\det(Y)N\epsilon}
\mbox{tr}(\tau_i h_I(v))\nonumber\\
E^{(\epsilon)a}_{i}(v) &:=& \frac{Y^a_I}{\det(Y)\epsilon^{D-1}} P^I_i(v)
\ea
Suppose first that $(h_I(v),P^I_i(v))\in G\times Lie(G)$ are obtained via 
the map $D_\epsilon$.
Then, using the smoothness of the fields $(A,E)$ it is easy to see
that $A^{(\epsilon)i}_a(v)-A_a^i(v)$ and 
$E^{(\epsilon)a}_i(v)-E^a_i(v)$ are both of order $\epsilon$.\\

We now select from the one-parameter family of lattices $\gamma_\epsilon$
and decompositions $P_{\gamma_\epsilon}$
a sequence of lattices $\gamma_n$ with the property that 
$\gamma_{n+1}$ and $P_{\gamma_{n+1}}$ respectively are {\it refinements} 
of $\gamma_n$ and $P_{\gamma_n}$ respectively. It is easy to see that
the following sequence does the job :
Start with some $\epsilon_0>0$ and call 
$\gamma_0:=\gamma_{\epsilon_0}, P_0:=P_{\gamma_{\epsilon_0}}$ 
respectively. Now consider the sequence $\epsilon_n:=\epsilon_0/3^n$.
The corresponding
$\gamma_n:=\gamma_{\epsilon_n}, P_n:=P_{\gamma_{\epsilon_n}}$ 
are obtained iteratively as follows :\\ 
Given $\gamma^0_{Rn}$, subdivide each of the cubes it defines into
$3^D$ axis parallel cubes of equal volume $(\epsilon_n/3)^D$ and similarly
for $P^0_{\gamma_{Rn}}$. Both of these lattices obviously refine the 
previous ones respectively. 

To see that the refinement of 
$P^0_{\gamma^0_{Rn}}$ has all the properties that we required it to have 
in the above construction of a decomposition of
$V_R$ dual to the refinement of $\gamma^0_{Rn}$ we remark the following : \\
If we consider $\gamma^0_{Rn}$ as a sublattice of an infinite regular cubic
lattice $L_n$ in $\Rl^D$ then $P^0_{\gamma^0_{Rn}}$ satisfies the required 
proprties if and only if its restriction to $C'_R$ is 
defined by a 
sublattice of the lattice $L'_n$ obtained from $L_n$ by translating it by the 
vector $\epsilon_n\sum_{I=1}^D b_I/2$. In other words, if we label the 
vertices of  $L_n$ by the n-tuples $(\epsilon_n n_I)_{I=1}^D,\;n_I\in Z$ then 
the vertices of $L'_n$ are labelled by the n-tuples 
$(\epsilon_n [n'_I+1/2])_{I=1}^D,\;n'_I\in Z$. Now the points of 
the refinements of 
$L_n$ and $L'_n$ respectively are labelled by 
$(\epsilon_n n_I/3)_I=\epsilon_{n+1} n_I$ and  
$(\epsilon_n [n'_I/3+1/2])_I=(\epsilon_n [(n'_I+1)/3+1/6])_I=
(\epsilon_{n+1}[n^{\prime\prime}_I+1/2])_I$, in other words, the refinements
coincide with $L_{n+1}$ and $L'_{n+1}$ respectively. Notice that
our procedure would work also if we would choose to refine by $k$ 
instead of $3$ where $k>3$ can be any odd integer.

To complete the decomposition we add cubes to these refinements as 
to fill $V_R$ as densely as possible according to the rules we specified 
above (also deleting cubes if necessary as discussed above) and thus 
obtain, after mapping with $X_R$, $\gamma_{n+1}$ and $P_{\gamma_{n+1}}$.   

We remark without proof that only cubic lattices seem to have the 
property that there are refinements such that a dual decomposition
of the refinement can be a refinement of the dual decomposition
(consider a simplicial decomposition to see the arising problems).\\
\\
Now that we know that if $v$ is a vertex of $\gamma_n$ for some $n$ 
then it is a vertex of all $\gamma_m$ for all $m\ge n$, the limit 
$\epsilon\to 0$ is well defined and we have, provided that
$A^{(\epsilon_n)i}_a(v)=:A^{(n)i}_a(v),\;
E^{(\epsilon_n)a}_i(v)=:E^{(n)a}_i(v)$ are defined via $D_n:=
D_{\epsilon_n}$
\ba \label{3.44}
&& \lim_{\epsilon\to 0}[A^{(\epsilon)i}_a(v)-A_a^i(v)] 
:=\lim_{n\to\infty} [A^{(n)i}_a(v)-A_a^i(v)]\nonumber\\
&& \lim_{\epsilon\to 0}[E^{(\epsilon)a}_i(v)-E^a_i(v)] 
:=\lim_{n\to\infty} [E^{(n)a}_a(v)-E^a_i(v)]
\ea
where convergence is pointwise on $M$. In other words, the map
$D_n:\; M\mapsto M_n\subset M_{\gamma_n}$ is invertible in the limit 
$n\to\infty$. To see that also the symplectic structure  $\Omega$
of $M$ is recovered in this limit we notice first that for each
$f_a^i,\; F^a_i\in {\cal S}$ we have 
$F(A)=\lim_{n\to\infty} F\cdot A^{(n)}(A), \;
E(f)=\lim_{n\to\infty} E^{(n)}(A,E)\cdot f$ pointwise in $M$
where 
\ba \label{3.45}
F\cdot A^{(n)}&:=& \sum_{v\in V(\gamma_n)}\epsilon_n^D
(\det(Y))(v) F_a^i(v) A^{(n)i}_a(v) \nonumber\\ 
E^{(n)}\cdot f&:=& \sum_{v\in V(\gamma_n)}\epsilon_n^D
(\det(Y))(v) E^{(n)a}_i(v) f_a^i(v) 
\ea
\begin{Theorem}\label{th3.4}
The bracket defined by
\ba 
\label{3.46a}
\{F(A),F'(A)\}'&:=&\lim_{n\to\infty} 
\{F\cdot A^{(n)}, F'\cdot A^{(n)}\}_{\gamma_n}\\     
\label{3.46b}
\{E(f),F(A)\}'&:=&\lim_{n\to\infty} 
\{E^{(n)}\cdot f, F\cdot A^{(n)}\}_{\gamma_n}\\     
\label{3.46c}
\{E(f),E(f')\}'&:=&\lim_{n\to\infty} 
\{E^{(n)}\cdot f, E^{(n)}\cdot f'\}_{\gamma_n}
\ea
for all $f_a^i,\; F^a_i,\; f_a^{\prime i},\; F^{\prime a}_i\in {\cal S}$ 
coincides with the symplectic structure $\Omega$ on $M$ defined 
by (\ref{3.13}).
\end{Theorem}
Proof of Theorem \ref{th3.4} :\\
We simply have to use the symplectic structure of 
$(M_{\gamma_n},\Omega_{\gamma_n})$ and take the limit using 
(\ref{3.44}). In the notation of this subsection, the symplectic
structure labelled by $\gamma_n$ can be written
\ba \label{3.47}
&& \{(h_I(v))_{AB},(h_J(v'))_{CD}\}_{\gamma_n}=0\nonumber\\
&& \{P^I_i(v),h_J(v')\}_{\gamma_n}=\delta^I_J\delta_{v,v'} 
\frac{\tau_i}{2}h_J(v)\nonumber\\
&& \{P^I_i(v),P^J_j(v')\}_{\gamma_n}=-\delta^{IJ}\delta_{vv'} 
f_{ij}\;^k P^I_k(v)
\ea
{[}1.]\\
Then (\ref{3.46a}) is obvious since the right hand side vanishes already 
at any finite $n$.\\
{[}2.]\\
We have at fixed $n$ for the right hand side of (\ref{3.46b})
\ba \label{3.48}
&&\{E^{(n)}\cdot f, F\cdot A^{(n)}\}_{\gamma_n}
= \sum_{v,v'\in V(\gamma_n)}\epsilon_n^{2D} 
(\det(Y))(v)(\det(Y))(v') f_a^i(v) F^b_j(v')\times\nonumber\\
&& \times [-2\frac{n_b^J(v')}{\det(Y)(v')N\epsilon_n}
\frac{Y^a_I(v)}{\det(Y)(v)\epsilon_n^{D-1}}]
\{P^I_i(v),\mbox{tr}(\tau_j h_J(v'))\}_{\gamma_n}
\nonumber\\
& =& -\frac{2}{N}\sum_{v\in V(\gamma_n)}\epsilon_n^D 
f_a^i(v) F^b_j(v)
[\sum_I (n_b^I Y^a_I)(v)\mbox{tr}(\tau_j \frac{\tau_i}{2} h_I(v))]
\nonumber\\
& =& -\frac{2}{N}\sum_{v\in V(\gamma_n)}\epsilon_n^D 
f_a^i(v) F^b_j(v)
[-\frac{N}{2}\delta_{ij} \delta^a_b\det(Y)(v)
+\sum_I (n_b^I Y^a_I)(v)\mbox{tr}(\tau_j \frac{\tau_i}{2} [h_I(v)-1])]
\nonumber\\
& =& \{\sum_{v\in V(\gamma_n)}\epsilon_n^D [\det(Y) f_a^i F^a_i](v)\}
\nonumber\\
&&
-\frac{2}{N}\{\sum_{v\in V(\gamma_n)}\epsilon^D [f_a^i F^b_j](v)
\sum_I (n_b^I Y^a_I)(v)\mbox{tr}(\tau_j \frac{\tau_i}{2} [h_I(v)-1])\}
\ea
Consider the first term in the last equality of (\ref{3.48}).  
We can write it as 
\be \label{3.49}
\sum_{R\in {\cal R}} \sum_{v^0\in \gamma^0_R} \epsilon_n^D 
|\det((\frac{\partial X}{\partial t}_{t(v^0)}| 
[f_a^i F^a_i](X(t(v^0)))
\ee
which defines a Riemann sum for the expression
\be \label{3.50}
\sum_{R\in {\cal R}} \int_{C_R} d^Dt
|(\det((\frac{\partial X}{\partial t})_{|t}| 
[f_a^i F^a_i](X(t))
=F(f)-\sum_{R\in {\cal R}} \int_{R-X_R(C_R)} d^Dx [f_a^i F^a_i](x)
\ee
and the second integral in (\ref{3.50}) vanishes in the limit 
$n\to \infty$ (that is $C_R\to R$). To see this, notice 
that in the limit $n\to \infty$
the integral $\int_{R-X_R(C_R)} d^Dx [f_a^i F^a_i](x)$ 
becomes $\int_{\partial R} d^D x [f_a^i F^a_i](x)$ which
vanishes for non-distributional spaces of test functions. 

Turning to the second term in (\ref{3.48}) we notice that 
there exists a positive constant $k$ such that 
$|\mbox{tr}(\tau_j \frac{\tau_i}{2} [h_I(v)-1])]|\le k\epsilon_n$
as $n\to\infty$ independent of the indices and $v$ because $A$ is 
a smooth and bounded function. Thus we see that the second term vanishes 
in the limit $n\to\infty$.\\
{[}3.]\\
Finally, at fixed $n$ the right hand side of (\ref{3.46c}) becomes
\ba \label{3.51}
\{E^{(n)}\cdot f, E^{(n)}\cdot f'\}_{\gamma_n}
&=& \sum_{v,v'\in V(\gamma_n)} \epsilon_n^2
[f_a^i Y^a_I](v) [f_b^{j\prime} Y^b_J](v') 
\{P^I_i(v),P^J_j(v')\}_{\gamma_n}\\
&=& -f_{ij}\;^k \sum_{v\in V(\gamma_n)} \epsilon_n^2
[\sum_I (f_a^i Y^a_I f_b^{j\prime} Y^b_I P^I_k)(v)]\nonumber\\
&=& -f_{ij}\;^k \epsilon_n\sum_{v\in V(\gamma_n)} \epsilon_n^D
[\sum_I (f_a^i Y^a_I f_b^{j\prime} Y^b_I [E^c_i n_c^I+
\{\frac{P^I_k}{\epsilon_n^{D-1}}-E^c_i n_c^I\}])(v)]
\nonumber
\ea
The whole sum is just an approximation for a Riemann integral times
$\epsilon_n$. 
The term in the curly bracket approaches zero as $n\to 0$ and is
therefore, together with the first term in the square bracket,
integrable against the product of test functions displayed.
Thus, the whole expression vanishes in the limit $n\to\infty$.\\
$\Box$\\

\subsection{Structured Graphs as Labels for Generalized Projective Families}
\label{s3.3}

A natural question to ask is whether one can identify $(M,\Omega)$ with
(the limit of) a generalized projective sequence of symplectic manifolds 
$(M_\gamma,\Omega_\gamma)$. The answer is affirmative but somewhat 
involved because we first must introduce new labels for projective families 
:

First of all, the family $(M_\gamma,\Omega_\gamma)$ does not only 
depend on the graph $\gamma$ but actually on the set $\cal L$ of 
{\it structured graphs} $l=(\gamma,P_\gamma,\Pi_\gamma)$ consisting of a 
graph $\gamma$,
a polyhedronal decomposition $P_\gamma$ dual to it and a choice of paths
$\rho_e(x)\in \Pi_\gamma$ adapted to $\gamma,P_\gamma$ where
$\rho_e(x)\subset S_e,\; e\in E(\gamma),\;x\in S_e$. The family $\cal L$
is partially ordered by inclusion but it is in general wrong that 
given two elements $l,l'\in{\cal L}$ there exists a common refinement,
that is, an element $\tilde{l}\in{\cal L}$ such that $l,l'\subset
\tilde{l}$. In other words, the inclusion relation does not equip $\cal L$
with the structure of a directed set on which the structure of a 
generalized projective limit crucially depends.

In order to proceed, we therefore must first modify the partial order. 
To motivate our choice we begin with the following observation :

Given a graph 
$\gamma$ the second and third entry of a structured graph $l$ such that
$\gamma(l)=\gamma$ are largely arbitrary. On the other hand,
if we consider structured graphs 
$l,l'$ with $\gamma(l)=\gamma(l')$ then by construction we 
can always find a diffeomorphism that preserves 
$\gamma(l)$ and maps $P_\gamma,\Pi_\gamma$ to $P_{\gamma'},\Pi_{\gamma'}$.
This follows from the fact that all the $S_e,\rho_e(x)\subset S_e,
x\in S_e$ are obtained via a diffeomorphism from a universal object
by definition \ref{def3.5}.
Now, while the actual {\it values} of the $P^e$ that we construct from
$l$ or $l'$ respectively may differ (the $h_e$ are evidently the same), the 
Poisson algebras, that is to say the {\it algebra of Hamiltonian vector 
fields}, that we obtain are identical. Moreover, let us consider the 
$h_e$ as elements of the space of smooth functions
$C^\infty({\cal C}_\gamma)$ 
of the configuration space ${\cal C}_\gamma$ of 
$M_\gamma$ (in fact they are coordinate functions)
and the $P^e$ as elements of the space of vector fields 
$V({\cal C}_\gamma)$ on ${\cal C}_\gamma$ via the map
$(h_e,P^e_j)\mapsto (h_e,\mbox{tr}((\tau_j h_e)^T\partial/\partial h_e)$.
The space $C^\infty({\cal C}_\gamma)\times V({\cal C}_\gamma)$ is equipped 
with the Lie algebra structure
$[(f,u),(f',u')]=(u(f')-u'(f),[u,u'])$ which is evidently closed 
and isomorphic with the Poisson bracket structure as obtained from both 
$l,l'$. In this form it is particularly obvious that both $l,l'$ give rise
to the same Lie algebra.

What this means is that the information contained in $l$ beyond that of
$\gamma(l)$ is irrelevant as far as the Poisson algebra is concerned.
Since it is the Poisson structure which sets the correspondence with 
quantum theory we will obtain isomorphic quantum theories from both $l,l'$.
The additional information contained in $l$ however comes in when we 
consider the classical limit of the theory. Namely, the coherent states 
constructed in \cite{46,47,45} are sensitive to the size and shape of the 
$S_e$ as well as the precise choice of the paths $\rho_e$.

These considerations shed light on the question why we have largely 
abused notation when writing $\gamma$ instead of $l$ and is also 
reflected in the subsequent definition. 
\begin{Definition} \label{def3.7}
We say that 
$(\gamma,P_\gamma,\Pi_\gamma)\prec(\gamma',P_{\gamma'},\Pi_{\gamma'})$
provided that $\gamma(l)\subset\gamma(l')$ and that $l,l'$ are 
equivalent, $l\equiv l'$, if $\gamma(l)=\gamma(l')$.
\end{Definition}
In other words, there are no
conditions at all on the second and third entry of a structured graph.
In particular, we identify $l$ with $l'$ if one obtains $l'$ form $l$ by 
applying a diffemorphism that preserves $\gamma(l)$. The subsequent
two lemmas are then almost trivial.
\begin{Lemma} \label{la3.2}
The relation $\prec$ defined in definition \ref{def3.7} defines a 
partial order.
\end{Lemma}
Proof of Lemma \ref{la3.2} :\\
1) Reflexivity : $l\prec l$ since $\gamma(l)=\gamma(l)$.\\
2) Antisymmetry : $l\prec l'$ and $l'\prec l$ implies $\gamma(l)=\gamma(l')$,
that is, $l\equiv l'$. \\
3) Transitivity : $l\prec l'$ and $l'\prec l^{\prime\prime}$ implies 
$\gamma(l)\subset \gamma(l')\subset\gamma(l^{\prime\prime})$, thus
$l\prec l^{\prime\prime}$.\\
$\Box$\\
\begin{Lemma} \label{la3.3}
The set $\cal L$ is directed, that is, for any given $l,l'\in{\cal L}$ there 
exists $\tilde{l}\in{\cal L}$ such that 
$l,l'\prec\tilde{l}$. Such an element $\tilde{l}$ is called a common
refinement of $l,l'$.
\end{Lemma}
Proof of Lemma \ref{la3.3} :\\
Given $l,l'\in {\cal L}$ consider the graph $\tilde{\gamma}:=
\gamma(l)\cup\gamma(l')$. Choose any $\tilde{l}\in {\cal L}$ such that 
$\tilde{\gamma}=\gamma(\tilde{l})$. Then $l,l'\prec\tilde{l}$.\\
$\Box$\\
Next we need the notion of a projection of symplectic manifolds. 
\begin{Definition} \label{def3.8}
Let $l\prec l'$, 
consider any edge $e\in E(\gamma)$ and find the edges
$e'_1,..,e'_n\in E(\gamma')$ such that $e=e'_1\circ..\circ e'_n$.
We then define the following projection
\be \label{3.53}
p_{l' l}:\; M_{l'}\mapsto M_l;
h_e:=h_{e'_1}..h_{e'_n} \mbox{ and } P^e:=P^{e'_1}
\ee
\end{Definition}
It is obvious that $p_{l' l}$ is onto for $l\prec l'$ except in 
the presence of boundary conditions in which case the $P^e_i$ for 
sufficiently small $S^e$ would be bounded. As a map between the 
$\bar{M}_l$ it would be onto. Define 
$m_l=\{h_e,P^e\}_{e\in E(\gamma)}$ and consider an array of non-singular
$\dim(G)\times\dim(G)$ matrices $\lambda=\{\lambda_e\}$ with an
action on the points of $M_l$ given by $\lambda\cdot m_l=
\{h_e,\lambda_e P^e\}$.
\begin{Definition} \label{def3.9}
i) Consider the uncountable direct product 
${\cal M}:=\times_{l\in {\cal L}} M_l$, then
the following subset 
\be \label{3.54}
{\cal M}_\infty:=\{(m_l)_{l\in {\cal L}}\in {\cal M};\;
\exists\;\lambda^{ll'}\;\ni\; \lambda^{ll'}\cdot m_l=
p_{l' l}(m_{l'})\;\forall\; l\prec l'\}
\ee
is called a generalized projective limit of the $M_l$.\\
ii) A family of symplectic structures $(\Omega_l)_{l\in{\cal L}}$
is called a self-consistent or generalized projective family provided that  
the associated Poisson brackets project in the usual way
\be \label{3.55}
p_{l' l}^\ast\{f,g\}_l:=\{p_{l' l}^\ast f, p_{l' l}^\ast g\}_{l'}
\; \forall\;f,g\in C^\infty(M_l)
\ee
that is, the $p_{l'l}$ are ``non-invertible'' symplectomorphisms.
\end{Definition}
It is easy to see that the symplectic structures $\Omega_l$ (or 
$\Omega_\gamma$ as we called them all the time) that we defined in 
section \ref{s3.2.2} form indeed a self-consistent family of 
symplectic structures on $M_l$ (or $M_\gamma$). This follows, as already
said,  
from the astonishing fact that the $\Omega_\gamma$
{\it are completely insensitive} to the size and shape of the faces 
of $P_\gamma$ and the choice of the paths of $\Pi_\gamma$ as long
as all the requirements of a dual decomposition are met. 
This is precisely the contents of the identities (\ref{3.21a}) -- 
(\ref{3.21c}).
This observation is tied to the fact that the smearing functions,
edges and faces, are sufficiently singular and that the smearing process
is background metric independent, so 
that only topological characteristics, such as intersection numbers of
edges with faces, are the results of the calculation and are thus 
completely shape independent, they are locally diffeomorphism invariant
(i.e. invariant under locally non-trivial diffeomorphisms).
Once more, this observation is also the logic behind definition \ref{def3.7} 
and 
behind labelling $(M_\gamma,\Omega_\gamma)$ only by elements of $\Gamma$ 
rather than by elements of $\cal L$.\\
\\
Let us then summarize :\\
We have shown in this subsection that the family of differentiable manifolds 
$(M_l)$ 
can be given the structure of a generalized projective limit 
${\cal M}_\infty$
and the family of symplectic structures thereon can be given
the structure of a self-consistent family of symplectic structures. 

In subsection \ref{s3.2.2} on the other hand we showed that there is a 
sequence $l_n\in {\cal L}$ (there denoted $\gamma_n$) such that 
$M=\lim_{n\to\infty} M_{l_n}$ and $\Omega=\lim_{n\to\infty} \Omega_{l_n}$
(pointwise limits). Moreover, $l_m\prec l_n$ for all $m\le n$, so
the sequence is linearly ordered. 
The points $m_{l_n}$ defined in section \ref{s3.2.2} belong to $M_{l_n}$.
By construction, we can extend every such sequence $m_{l_n}$ 
to a sequence $(m_l)_{l\in {\cal L}}\in{\cal M}_\infty$. (Explicitly, 
the array of matrices is for $n>m\to\infty$ approximately given by 
$\lambda_{eij}^{l_m l_n}=\delta_{ij} (\epsilon_n/\epsilon_m)^{D-1}$).
It follows that the sequence $(m_{l_n})$ can be embedded into a
generalized projective sequence which in turn defines 
a point of ${\cal M}_\infty$. Likewise, 
the standard symplectic manifold $(M,\Omega)$ 
can be identified with the sequence $(M_{l_n},\Omega_{l_n})$  
of symplectic manifolds which in turn can be extended to
a subset of the generalized projective limit and self-consistent symplectic 
structures thereon respectively (with respect to the generalized projective 
limit ${\cal M}_\infty$).\\
\\
Remark :\\
Of course, we have displayed $(M,\Omega)$ only as the union of a 
{\it very special} subset of {\it all} generalized projective sequences. 
An arbitrary generalized projective sequence
wil not have any obvious interpretation 
in terms of connections and electric fields on {\it any smooth manifold}
$\Sigma$. This is possible because 
the set of graphs in -- and dual decompositions of $\Sigma$ have much more 
structure than the set of points of $\Sigma$. In fact, the picture 
that emerges is {\it completely combinatorical} and only very special 
configurations of graphs and dual decomposition allow a manifold
interpretation. In a sense, without specifying the embedding 
of abstract graphs and dual decompositions into a concrete $\Sigma$ 
we are treating all manifolds $\Sigma$ {\it simultaneously}. Thus,
although in the canonical approach to quantum gravity one starts with a 
given differential manifold, the emerging classical and quantum theory
does not depend any longer on the particular choice of $\Sigma$. 
Only if one insists on a manifold interpretation there will be restrictions
on possible graphs (they have to agree, for instance with the Euler
characteristic of $\Sigma$ and the dimension of $\Sigma$) and on the 
spectra of operators 
\cite{13}. This opens the possibility to describe topology change within
canonical quantum gravity.

\section{The Gauss Constraint}
\label{s4}

In this section we implement the Gauss constraint into the theory.
On $(M,\Omega)$ it is given by the function ($\Lambda^i\in {\cal S}$)
\be \label{4.1}
G(\Lambda):=\int_\Sigma d^Dx \Lambda^i(x)[\partial_a E^a_i(x)+
f_{ij}\;^k A_a^j(x) E^a_k(x)]
\ee
which generates infinitesimal gauge transformations
\ba \label{4.2}
F(A) &\mapsto& F(A)+\{F(A),G(\Lambda)\}=F(A+D\Lambda) \nonumber\\
E(f) &\mapsto& E(f)+\{E(f),G(\Lambda)\}=E(f+[\Lambda,f])
\ea
where $\Lambda=\Lambda^i\tau_i/2,A=A^i\tau_i/2,E=E_i\tau_i/2,f=f^i\tau_i/2,
F=F_i\tau_i/2$. The maps (\ref{4.2}) are the infinitesimal versions of 
the finite gauge transformations
\ba \label{4.3}
F(A) &\mapsto& F(\mbox{Ad}_g\cdot A-dg g^{-1}) \nonumber\\
E(f) &\mapsto& [\mbox{Ad}_g\cdot E](f)
\ea
where $\mbox{Ad}_g\cdot v:=g vg^{-1}$ is the adjoint representation of 
$G$ on $Lie(G)$. Indeed, for infinitesimal $\Lambda$, (\ref{4.3}) 
reproduces (\ref{4.2}) to linear order provided we identify
$g(x)=\exp(-\Lambda^i(x)\tau_i/2)$.

Our task is to write (\ref{4.1}) in terms of $(M_\gamma,\Omega_\gamma)$.
That is, we must find a function $G_\gamma(\Lambda)$ on $M_\gamma$
such that it converges pointwise on $M$ to $G(\Lambda)$ and such that
the limit of its Poisson brackets, that is, its Hamiltonian vector field
on $M_\gamma$ converges to the Hamiltonian vector field of $G(\Lambda)$
on $M$.

To do this we will proceed as follows :\\
1) Find the gauge transformations of the coordinates $h_e,P^e$ of
$M_\gamma$ induced by (\ref{4.2}).\\
2) Find a generator $G_\gamma(\Lambda)$ on $M_\gamma$ of these infinitesimal
transformations.\\
3) Study the generator and its Hamiltonian vector field on $M_\gamma$ and 
consider the limit $\gamma\to\Sigma$.\\
Notice that this procedure works only because gauge transformations have 
the special feature to preserve $M_\gamma$ as we will see. This is not 
the case for more general gauge groups such as diffeomorphisms which
map between different $M_\gamma$'s and which are relevant for
quantum general relativity \cite{23f}.

It is immediate from the definition of a principal fibre bundle with 
connection over $\Sigma$ and an associated (under the adjoint 
representation of $G$) vector bundle that under finite gauge transformations
$x\in\Sigma\mapsto g(x)\in G$
\ba \label{4.4}
h_e(A)&\mapsto& g(e(0)) h_e(A) g(e(1))^{-1} \nonumber\\
P^e(A,E)&\mapsto& g(e(0)) P^e(A,E) g(e(0))^{-1}
=:\mbox{Ad}_{g(e(0))}\cdot P^e(A,E)
\ea
where $P^e=P^e_i\tau_i/2$. This follows from the manifestly gauge 
covariant definition of the basic coordinates of $M_\gamma$
given in (\ref{3.14}). The infinitesimal version of (\ref{4.4})
is given by (with $g(x)=\exp(-\Lambda^i(x)\tau_i/2)=\exp(-\Lambda(x))$)
\ba \label{4.5}
h_e&\mapsto& h_e-\Lambda(e(0))h_e-h_e \Lambda(e(1)) \nonumber\\
P^e &\mapsto& P^e +[P^e,\Lambda(e(0))]
\ea
which should equal 
$\{h_e,G_\gamma(\Lambda\}_\gamma,\{P^e,G_\gamma(\Lambda)\}_\gamma$
respectively.

It is immediately clear from (\ref{3.21c}) that the second line of 
(\ref{4.5}) can be obtained by choosing
\be \label{4.6}
G_\gamma(\Lambda)=\sum_{v\in V(\gamma)}\Lambda^i(v) \sum_{e\in 
E(\gamma),e(0)=v} P^e_i +\mbox{ more}
\ee
where ``more'' should commute with all the $P^e_i$. Ansatz (\ref{4.6})
already correctly reproduces also the $\Lambda(e(0))$
term of the first line of (\ref{4.5}), the $\Lambda(e(1))$ term looks 
similar just that it corresponds to an insertion of $\tau_i$ from the 
right instead of from the left. Since the holonomies Poisson commute
among themselves we are led to the following improved ansatz
\be \label{4.7}
G_\gamma(\Lambda)=\sum_{v\in V(\gamma)}\Lambda^i(v) 
[\sum_{e\in E(\gamma),e(0)=v} P^e_i+\sum_{e\in E(\gamma),e(1)=v} 
M_{ij}(h_e) P^e_j] 
\ee
where the matrix $M_{ij}(h_e)$ should satisfy
$-M_{ij}(h_e)\tau_j h_e=h_e \tau_i$ and 
$\{P^e_i,M_{jk}(h_e) P^e_k\}_\gamma=0$. The first requirement
leads to the unique solution 
\be \label{4.8}
M_{ij}(h)=\frac{1}{N}\mbox{tr}(h\tau_i h^{-1}\tau_j)
\ee
while the second asks us to check the vanishing of 
(use $\{.,h^{-1}\}_G=-h^{-1}\{.,h\}_G h^{-1}$)
\ba \label{4.9}
\{P_i,M_{jk}(h) P_k\}_G 
&=& -M_{jk}(h)f_{ik}\;^l P_l+\{P_i,M_{jl}(h)\}_G P_l\nonumber\\
&=& [-M_{jk}(h)f_{ik}\;^l +\frac{1}{N}
\mbox{tr}(\frac{\tau_i}{2}h\tau_j h^{-1}\tau_l 
-h\tau_j h^{-1}\frac{\tau_i}{2}\tau_l)] P_l\nonumber\\
&=& [-f_{ik}\;^l + f_{li}\;^k]M_{jk}(h) P_l
\ea
which indeed vanishes for $G$ semisimple as we assume.

We notice that 
\be \label{4.10}
P^{e\prime}_i:=-M_{ij}(h_e) P^e_j=P^{e^{-1}}_i
\ee
which explains intuitively why it is possible that 
$\{P^e_i,P^{e'\prime}_j\}_\gamma$ vanishes for any $e'$ : while 
$P^e$ depends only on the beginning half segment of the edge $e$, 
$P^{e\prime}$ depends only on the ending half segment of the edge $e$
and given the symplectic structure (\ref{3.13}) a non-vanishing
bracket is therefore impossible (modulo the regularization procedure
of section \ref{s3.2.1}). Of course, in retrospect the result should
have been guessed on general grounds as what we were trying to construct
are the generators $P,P'$ respectively of left and right translations 
respectively on $G$ which, of course, commute.\\
\\
In order to check the continuum limit of the function
(\ref{4.7}) we employ the sequence of graphs $\gamma_n$ of section
\ref{s3.2.2}. Using the notation of that section, in particular
(\ref{3.42a}), we 
define for $v\in \gamma_R,\;R\in {\cal R}$ the quantity 
$$
E^I(v)=
h_{e_I(v)}(0,1/2)^{-1}[\int_{S^I(v)} h_{\rho_{e_I(v)}(x)} 
\ast E(x) h_{\rho_{e_I(v)}(x)}^{-1}] h_{e_I(v)}(0,1/2)
$$
which to order $\epsilon_n^{D-1}$ equals $\epsilon_n^{D-1}n_a^I(v)
E^a_i(v)\tau_i$ as $n\to\infty$. Then for fixed $n$
\ba \label{4.11}
&& G_{\gamma_n}(\Lambda)
= \sum_{R\in {\cal R}}\sum_{v\in V(\gamma_R)}\Lambda^i(v) 
[\sum_{e\in E(\gamma_R),v=e(0)} P^e_i
+\sum_{e\in E(\gamma_R),v=e(1)} M_{ij}(h_e)P^e_j]
\nonumber\\
&=& \sum_{R\in {\cal R}}\sum_{v\in V(\gamma_R)}\sum_{I=1}^D
[P^I_i(v)\Lambda^i(v)+M_{ij}(h_I(v)) 
P^I_j(v)\Lambda^i(X_R(X^{-1}_R(v)+\epsilon b_I))] \nonumber\\
\nonumber\\
&=& \sum_{R\in {\cal R}}\sum_{v\in V(\gamma_R)} \Lambda^i(v)\sum_{I=1}^D
[P^I_i(v)+M_{ij}(h_I(X_R(X^{-1}_R(v)-\epsilon b_I))) 
P^I_j(X_R(X^{-1}_R(v)-\epsilon b_I))] \nonumber\\
&=& -\frac{1}{N}\sum_{R\in {\cal R}}\sum_{v\in V(\gamma_R)} 
\Lambda^i(v)\sum_{I=1}^D
\mbox{tr}(\tau_i[h_I(v)E^I(v)h_I(v)^{-1}
-E^I(X_R(X^{-1}_R(v)-\epsilon b_I))] \nonumber\\
&=& -\frac{1}{N}\sum_{R\in {\cal R}}\sum_{v\in V(\gamma_R)} 
\Lambda^i(v)\sum_{I=1}^D
\mbox{tr}(\tau_i[\{h_I(v)E^I(v)h_I(v)^{-1}-E^I(v)\}
\nonumber\\
&& +\{E^I(v)-E^I(X_R(X^{-1}_R(v)-\epsilon b_I))\}]) 
\ea
Consider the two curly brackets in the last line of (\ref{4.11}).
The first one is given to leading order $\epsilon^D$ by
$Y^b_I(v) n_a^I(v)[A_b(v),E^a(v)]=\det(Y_R)(v)[A_a(v),E^a(v)]$. The second
one is given to leading order $\epsilon_n^{D}$ by 
$$
[\frac{\partial(n_a^I(X_R(t)) E^a(X_R(t)))}{\partial t^I}]_{X_R(t)=v}
=n_a^I(v) Y^b_I(v)\partial_b E^a(v)=\det(Y_R)(v)\partial_a E^a(v)
$$
since $\sum_I \partial_I n_a^I(X_R(t))=0$. Now the sum of the differences\\
$h_I(v)E^I(v)h_I(v)^{-1}-E^I(v)-\epsilon_n^D\det(Y_R)(v)[A_a(v),E^a(v)]$
and\\  
$E^I(v)-E^I(X_R(X^{-1}_R(v)-\epsilon b_I))
-\epsilon_n^D\det(Y_R)(v)\partial_a E^a(v)$
can be written as $\epsilon_n^{D+1} \det(Y_R)(v)K(v)$ where $K$ is an 
integrable function. Thus, (\ref{4.11}) becomes
\ba \label{4.12}
G_{\gamma_n}(\Lambda)
&=& \sum_{R\in {\cal R}}\sum_{v\in V(\gamma_R)}\epsilon_n^D\det(Y_R)(v)
\Lambda^i(v)(\partial_a E^a_i+f_{ij}\;^k A_a^j E^a_k)(v)\nonumber\\
&& -\frac{\epsilon_n}{N}\sum_{R\in {\cal R}}\sum_{v\in V(\gamma_R)}
\epsilon_n^D
\det(Y_R)(v)\Lambda^i(v)\sum_{I=1}^D \mbox{tr}(\tau_i K(v))
\ea
and both sums are Riemann sum approximations of integrals. Recall that
$\det(Y_R)(v)>0$ for $v\in R$ and that $\epsilon_n^D\det(Y)(v)$ approximates
the Lebesgue measure of the image under $X_R$ of a cube of volume $\epsilon$
in $V_R$. It follows that 
\be \label{4.13}
\lim_{n\to\infty} G_{\gamma_n}(\Lambda)= G(\Lambda)
-\lim_{n\to\infty} \frac{\epsilon_n}{N}
\int_\Sigma d^Dx \Lambda^i(x)\mbox{tr}(\tau_i K(x))
=G(\Lambda)
\ee
as desired.\\
\\
To check that also the Hamiltonian vector field of $G_\gamma(\Lambda)$
converges to the one of $G(\Lambda)$ we consider the brackets defined
by
\ba \label{4.14}
\{F(A),G(\Lambda)\}'&:=&\lim_{n\to \infty}
\{F\cdot A^{(n)},G_{\gamma_n}(\Lambda)\}_{\gamma_n}
\nonumber\\
\{E(f),G(\Lambda)\}'&:=&\lim_{n\to \infty}
\{E^{(n)} \cdot f,G_{\gamma_n}(\Lambda)\}_{\gamma_n}
\ea
where $A^{(n)},E^{(n)}$ are defined in (\ref{3.44}), (\ref{3.45}).
Using the definitions and reasonings repeatedly outlined already in this 
paper we see that indeed
$\{F(A),G(\Lambda)\}'=\{F(A),G(\Lambda)\}$ and 
$\{E(f),G(\Lambda)\}'=\{E(f),G(\Lambda)\}$. So, the Hamiltonian vector fields
also coincide in the limit $\gamma\to \Sigma$ and display (\ref{4.7})
as a satisfactory discretization of $G(\Lambda)$.

\section{Quantization}  
\label{s5}

In this section we quantize all the phase spaces 
$(M_\gamma,\Omega_\gamma)$.
Notice that in the literature so far one quantized either $(M,\Omega)$
\cite{7} or one quantized only one particular family of 
$(M_\gamma,\Omega_\gamma)$'s that were defined through lattices  
in $\Sigma$'s of the topology of $\Rl^3$ \cite{42}.
In the former case one took a classical function and tried to turn it
into an operator after going through some regularization and 
renormalization steps. In the latter case one started directly with
some operators and required that they have a certain continuum limit 
behaviour with respect to the lattice spacing, however, one did not
establish a precise relation between these discrete operators and 
certin smeared objects of the continuum theory as we did in section 
\ref{s3}. However, without such an analysis it is quite unclear what the 
operators so obtained do actually measure. In particular,
one has to postulate the $\epsilon$ expansion of the $P^e,h_E$ rather 
than being able to derive it from first principles.\\
\\
By definition, quantization means to find an irreducible representation
of an algebra of operators $\hat{h}_e,\hat{P}_e$ such that the symplectic 
and the reality structure of the classical theory is correctly implemented.
More concretely, since $M_\gamma$ is isomorphic with the direct product
of co-tangent bundles $T^\ast G$, one copy for each edge of $\gamma$,
it is suggested to choose the natural real polarization of the phase space
in which wave functions depend only on holonomies.
Thus we choose a Hilbert space ${\cal H}_\gamma$ of square 
integrable functions of the $h_e,\; e\in E(\gamma)$ with respect to
a measure $\mu_\gamma$, that is, 
${\cal H}_\gamma=L_2({\cal C}_\gamma,d\mu_\gamma)$ 
where ${\cal C}_\gamma=G^{|E(\gamma)|}$ is the complete quantum (and 
also classical in the absence of boundary conditions) configuration space
and must represent 
the operators $\hat{h}_e^{AB},\hat{P}^e_i$ on ${\cal H}_\gamma$ in such a way
that the following commutation relations hold :
\ba \label{5.1}
{[}\hat{h}_e^{AB},\hat{h}_{e'}^{CD}] &=& 0\nonumber\\
{[}\hat{P}^e_j,\hat{h}_{e'}^{AB}] &=&  i\hbar 
\delta^e_{e'}(\frac{\tau_j}{2}\hat{h}_e)^{AB}\nonumber\\
{[}\hat{P}^e_j,\hat{P}^{e'}_k] &=&  i\hbar 
\delta^{ee'}(-f_{jk}\;^l) \hat{P}^e_l
\ea
More precisely, we must find a common dense domain ${\cal D}_\gamma$
of all the basic operators which they leave invariant so that it makes sense
to compute commutators. Notice again that we allow the graph $\gamma$ to 
be infinite. 

Furthermore, the reality structure of the classical theory is given by
(let us choose $G$ to be a (subgroup of a) unitary group for definiteness)
\be \label{5.2}
\overline{h_e^{AB}}=(h_e^{-1})^{BA} \mbox{ and }
\overline{P^e_i}=P^e_i
\ee
To see the latter, notice that from (\ref{3.14}) $P^e_i$ is given by an
integral of quantities of the form $\mbox{tr}(g\tau_i g^{-1}\tau_j) v^j$
where $v^j$ is real and $g\in G$. From $\bar{g}^T g=1$ it follows with
$g=\exp(\theta^j\tau_j/2),\;\theta^j$ real that $\bar{\tau}^T_j=-\tau_j$.
Therefore the orthogonal matrix, using $\mbox{tr}(M^T)=\mbox{tr}(M)$,
\be \label{5.3} 
O_{ij}(g):=-\frac{1}{N}\mbox{tr}(g\tau_i g^{-1}\tau_j)
\ee
is real. 

In conclusion we must impose the following adjointness relations on 
$\mu_\gamma$ 
\be \label{5.4}
(\hat{h}_e^{AB})^\dagger=\widehat{(h_e^{-1})^{BA}} \mbox{ and }
(\hat{P}^e_i)^\dagger=\hat{P}^e_i
\ee
where the first identity has to be understood in the sense that one should 
write the function $h_e^{-1}$ in terms of $h_e$ and then replace it by
$\hat{h}_e$. No operator ordering problems arise since the $h_e^{AB}$ are 
mutually commuting.
As advertized, we will choose the $\hat{h}_e^{AB}$ as multiplication 
operators with values
in $G$. As $G$ is compact, these operators are bounded and thus they are
defined, together with $\widehat{(\hat{h}_e^{-1})^{AB}}$, everywhere
on ${\cal H}_\gamma$ so that there are no domain questions at all
in the definition of $(\hat{h}_e^{AB})^\dagger$. The second identity in
(\ref{5.4}) says that $\hat{P}^e_j$ is a self-adjoint operator and in 
order to
settle the domain question we will 
determinine an explicit core ${\cal D}_\gamma$ of
essential self-adjointness for all the $\hat{P}^e_i$.

Let us choose as ${\cal D}_\gamma:=C^\infty({\cal C}_\gamma)$ where we 
consider
${\cal C}_\gamma$ as a Banach manifold modelled on $\Rl^{\dim(G)|E(\gamma)|}$
similar as for $M_\gamma$. Then we choose the following action
of the basic operators on $f_\gamma\in {\cal D}_\gamma$  
\ba \label{5.5}
(\hat{h}_e^{AB} f_\gamma)(\{h_{e'}\})&:=&
h_e^{AB} f_\gamma(\{h_{e'}\}) \nonumber\\
(\hat{P}^e_j f_\gamma)(\{h_{e'}\})&:=&
\frac{i\hbar}{2} (X^e_j f_\gamma)(\{h_{e'}\}) 
\ea
where $X^e_j=X(h_e)_j,\;X(g)_i:=\mbox{tr}((\tau_j g)^T\partial/\partial g)$
denotes the right invariant vector field on $G$ (the generator of left 
translations). First of all, the operations (\ref{5.5}) leave ${\cal 
D}_\gamma$ obviously invariant. Next we have the Lie algebra of vector 
fields on $G$ given by $[X_j,X_k]=-2 f_{jk}\;^l X_l$ and it is easy to see
that with this choice the commutation relations (\ref{5.1}) are 
identically satisfied.

The direct product structure of ${\cal C}_\gamma$
shows that we may choose 
$$
d\mu_\gamma(\{h_e\}_{e\in E(\gamma)})=\otimes_{e\in E(\gamma)}
d\mu_e(h_e)
$$
and in order that the adjointness relations (\ref{5.4}) be satisfied it 
will be sufficient to choose $\mu_e=\mu_G$ for all $e\in E(\gamma)$.
Let ${\cal H}_G=L_2(G,d\mu_G)$, then we must establish
the symmetry property 
\be \label{5.6}
<f,i X_j f'>_G=i\int_G d\mu_G(h) \overline{f(h)} (X_j f')(h)
=<i X_j f,f'>_G
\ee
for all $f,f'\in C^\infty(G)$. Notice that ${\cal D}(X^\dagger)$,
the set of elements $f\in{\cal H}_G$ for which the map
$\psi\mapsto <f,iX_j \psi>$ defines a continuous linear functional on 
${\cal D}(X)=C^\infty(G)$, certainly contains ${\cal D}(X)$ so that  
(\ref{5.6}) implies symmetry.\\
Now $X_j$ generates left translations and
$\partial G=\emptyset$, thus, if we choose the measure $\mu_G$ to be left
invariant we are done provided we can establish that the expression for $X_j$
is real valued. For compact groups the only solution is, up to a 
normalization, $\mu_G=\mu_H$, the Haar measure on $G$ which is 
simultaneously left and right invariant and normalized, $<1,1>_G=1$.
To see that the expression for $X_j$ is real we perform the following 
calculation :
\ba \label{5.7} 
<f,X_j f'>_G &=& \int_G d\mu_H(h) \overline{f(h)} 
\frac{d}{dt}_{t=0} f'(e^{t\tau_j/2}h) \nonumber\\
&=& 
\frac{d}{dt}_{t=0}
\int_G d\mu_H(h) \overline{f(h)} f'(e^{t\tau_j/2}h) \nonumber\\
&=& 
\frac{d}{dt}_{t=0}
\int_G d\mu_H(e^{-t\tau_j/2}h) \overline{f(e^{-t\tau_j/2}h)} f'(h) 
\nonumber\\
&=& 
-\frac{d}{dt}_{t=0}
\int_G d\mu_H(h) \overline{f(e^{t\tau_j/2}h)} f'(h) 
\nonumber\\
&=& -<X_j f,f'>_G
\ea
as claimed.\\
Finally, to see that $i X_j$ is essentially self-adjoint with core
${\cal D}(X)=C^\infty(G)$ we show that 
$[X_j\pm \mbox{id}_{{\cal H}_G}]{\cal D}_G$ is dense
in ${\cal H}_G$ (basic criterion of essential self-adjointness). 
The proof is simplified through an appeal to the Peter\&Weyl 
theorem \cite{43} : the Hilbert space ${\cal H}_G$ is the completion 
of a countable
orthogonal sum of finite dimensional Hilbert spaces ${\cal H}_\pi$
where $\pi$ runs through the set of equivalence classes of irreducible
representations of $G$. A complete orthonormal basis of ${\cal H}_\pi$
is given by the functions $\sqrt{d_\pi}\pi_{mn}(h)$ where $d_\pi$ 
is the dimension of the representation and $\pi_{mm'}$ denotes the 
matrix elements of an abrbitary but fixed representant of that equivalence
class. These functions obviously belong to ${\cal D}(X)$ and
finite linear combinations of such functions are still
in ${\cal D}(X)$. Thus, the finite linear combinations of such 
functions belong to the domain, $\oplus_\pi {\cal H}_\pi 
\subset {\cal D}(X)$.

Next, it is easy to see that $X_j$ preserves ${\cal H}_\pi$. Denote by 
$X^\pi_j$ the restriction of $X_j$ to ${\cal H}_\pi$ then $i X^\pi_j$
is a symmetric operator on the finite dimensional Hilbert space 
${\cal H}_\pi$ and therefore self-adjoint on ${\cal H}_\pi$ with
domain given by all of ${\cal H}_\pi$. By the basic criterion for 
self-adjointness, $[X^\pi_j\pm 1]{\cal H}_\pi={\cal H}_\pi$. The proof 
is then complete with the observation that
\ba \label{5.8}
&&[X_j\pm 1]\oplus_\pi {\cal H}_\pi=
\oplus_\pi ([X_j^\pi\pm 1]{\cal H}_\pi)=\oplus_\pi {\cal H}_\pi
\subset[X_j\pm 1]{\cal D}(X) \nonumber\\
&\Rightarrow&
{\cal H}_G=\overline{\oplus_\pi {\cal H}_\pi}
\subset \overline{[X_j\pm 1]{\cal D}(X)}\subset {\cal H}_G 
\ea
Remark :\\
To see that $X^\pi_j$ does not have real eigenvectors in a more elementary 
way, recall that
$(X^\pi_j)^2=-\lambda_\pi<0$ is the Laplacian on $G$.\\
\\
In conclusion the (possibly infinite) tensor product of Hilbert spaces
\be \label{5.8a}
{\cal H}_\gamma:=\otimes_{e\in E(\gamma)} {\cal H}_e=L_2({\cal C}_\gamma,
d\mu_{0\gamma}=\otimes_{e\in E(\gamma)} d\mu_e)
\ee
where each of the ${\cal H}_e$ is isomorphic with $L_2(G,d\mu_H)$
is a faithful representation of the canonical commutation relations 
(\ref{5.1}) and of the adjointness relations (\ref{5.4}). Moreover,
given the action (\ref{5.5}), it is easy to see that the product Haar 
measure $\mu_{0\gamma}$ is the unique solution, that is, any other measure 
$\mu_\gamma$ which is regular with respect to it must be a constant
multiple of it. Notice that infinite products of probability measures
are well-defined and $\sigma-$additive probability measures by the  
Kolmogorov theorem \cite{44}. Much more will be said about infinite 
tensor products of Hilbert spaces in the first reference of \cite{45}.\\
\\
We now must quantize various functions on $M_\gamma$. It is at this point
where our detailed analysis becomes crucial : while in the continuum theory
important functions such as the Gauss constraint (\ref{4.1}) are written
as integrals over polynomials of the field variables $A(x),E(x)$ 
{\it at the same point}, that is, not as polynomials of the smeared
functions $F(A),E(f)$, the functions on $M_\gamma$ are polynomials 
of the basic observables $h_e,P^e$ {\it which are already smeared}.
Thus, while the quantization of, say, $G(\Lambda)$ on $(M,\Omega)$ can 
possibly produce
UV divergent objects, the quantization of $G_\gamma(\Lambda)$ 
on $(M_\gamma,\Omega_\gamma)$
{\it cannot suffer from such problems}. (Of course, in both cases factor 
ordering problems
might appear but in the latter case this is only an ambiguity and not
the source of a divergence). One might think that problems occur when 
taking the limit $\gamma\to\Sigma$, but as we will show, this does not 
happen. 

On the other hand, in both cases we can still
have IR divergencies. However, again, from our point of view this is {\it 
not a problem at all !} Namely, our operator {\it is densely defined}, that
is, it is an unbounded operator defined on a dense domain. This dense subset
of the Hilbert space, however, does not contain states with infinite volume.
Nevertheless it is possible to deal with this situation appropriately
\cite{45}. In contrast, the perturbative quantization of general relativity
is based on a cyclic vector with infinite volume and therefore IR 
divergencies necessarily occur.

Let us then quantize the Gauss constraint $G_\gamma(\Lambda)$. 
We choose to order the momentum variables to the right of the 
configuration variables and obtain
\be \label{5.9}
\hat{G}_\gamma(\Lambda)=
\sum_{e\in E(\gamma)} [\Lambda^i(e(0))\delta_{ij}
-\Lambda^i(e(1)) O_{ij}(\hat{h}_e)]\hat{P}^e_j
\ee
Let us check that there are no quantum anomalies. First of all
we compute the classical constraint algebra. We have
\ba \label{5.10}
\{O_{ij}(h) P_j,O_{kl}(h) P_l\}_G &=&
\{O_{ij}(h),O_{kl}(h) P_l\}_G P_j=
O_{kl}(h)\{O_{ij}(h),P_l\}_G P_j\\ 
&=& \frac{1}{N} O_{kl}(h)
\mbox{tr}(\frac{\tau_l}{2}h\tau_i h^{-1} \tau_j
-h\tau_i h^{-1} \frac{\tau_l}{2}\tau_j) P_j
=- f_{jl}\;^m O_{kl}(h) O_{im}(h)
P_j\nonumber
\ea
Now since 
\be \label{5.11}
h\tau_i h^{-1}=\mbox{Ad}_h\cdot\tau_i=O_{ij}(h)\tau_j
\ee
we have the identity
\ba \label{5.12}
&& [\mbox{Ad}_h\cdot\tau_i,\mbox{Ad}_h\cdot\tau_j]
=O_{ik}(h)O_{jl}(h)[\tau_k,\tau_l]
=\mbox{Ad}_h\cdot[\tau_i,\tau_j]
\nonumber\\
&\Rightarrow& 
O_{ik}(h)O_{jl}(h) f_{kl}\;^m=f_{ij}\;^k O_{km}(h)
\ea
which, when inserted into (\ref{5.10}), gives 
\be \label{5.13}
\{O_{ij}(h) P_j,O_{kl}(h) P_l\}_G
=- f_{lm}\;^j O_{kl}(h) O_{im}(h)P_j=
f_{ik}\;^m O_{mj}(h)P_j
\ee
or, recalling (\ref{4.10}),
\be \label{5.14}
\{P^{e\prime}_i,P^{e'\prime}_j\}_\gamma
=\delta^{e e'} f_{ij}\;^k P^{e\prime}_k
\ee
which is the algebra of {\it left invariant} vector fields on $G$ (notice 
the relative minus sign as compared to (\ref{3.21c}).\\ 
Thus, since the $P^e_i,P^{e'\prime}_j$ Poisson commute by definition
of the matrix $O_{ij}=-M_{ij}$ we immediately get with
\be \label{5.15}
G_\gamma(\Lambda)=\sum_{v\in V(\gamma)}\Lambda^i(v)
[\sum_{e(0)=v}P^e_i-\sum_{e(1)=v}P^{e\prime}_i]
\ee
that
\be \label{5.16}
\{G_\gamma(\Lambda),G_\gamma(\Lambda')\}_\gamma =
\sum_{v\in V(\gamma)} \Lambda^i(v)\Lambda^{j\prime}(v)f_{ij}\;^k
[-\sum_{e(0)=v}P^e_k+\sum_{e(1)=v} P^{e\prime}_k]
=-G_\gamma([\Lambda,\Lambda'])
\ee
as desired because we infer from (\ref{4.2}) that the continuum 
Poisson algebra is given by (use the Jacobi identity)
\ba \label{5.17}
&&\{\{E(f),G(\Lambda)\},G(\Lambda')\}
-\{\{E(f),G(\Lambda')\},G(\Lambda)\}
=\{E(f),\{G(\Lambda),G(\Lambda')\}\}
\nonumber\\
&=& 
E([\Lambda',[\Lambda,f]]-[\Lambda,[\Lambda',f]])
=E([f,[\Lambda',\Lambda]])
=-\{E(f),G([\Lambda,\Lambda'])\}
\ea
Thus, the algebra (\ref{5.16}) converges to (\ref{5.17}) by (\ref{4.13}).

Now, it follows trivially from (\ref{5.1}), (\ref{5.9}) that 
\be \label{5.18}
[\hat{G}_\gamma(\Lambda),\hat{G}_\gamma(\Lambda')]=i\hbar
(-\hat{G}_\gamma([\Lambda,\Lambda'])) 
\ee
as required.\\
\\
Let us summarize :\\
We started with a continuum phase space $(M,\Omega)$ and derived from it
a discrete phase space $(M_\gamma,\Omega_\gamma)$ for every graph $\gamma$.  
We also showed that $(M,\Omega)$ is the pointwise limit of 
$(M_\gamma,\Omega_\gamma)$ as $\gamma\to\Sigma$.
Next, we took the Gauss constraint $G(\Lambda)$ which is a function on
$M$ and derived from it a function $G_\gamma(\Lambda)$ on $M_\gamma$ which
again converges pointwise to $G(\Lambda)$. Moreover, the Poisson algebra
of the $G_\gamma(\Lambda)$ with respect to $\Omega_\gamma$ closes for 
every fixed $\gamma$ and converges pointwise to the Poisson algebra of the
$G(\Lambda)$. Finally, we quantized the $G_\gamma(\Lambda)$ and obtained
an anomaly free algebra of quantum constraints $\hat{G}_\gamma(\Lambda)$.
Then two questions remain : \\
1.) Does this structure provide us with 
a quantization of $G(\Lambda)$ as well ? That is, can we find an operator
$\hat{G}(\Lambda)$ densely defined on all of $\cal H$ and not only
on ${\cal H}_\gamma$ such that
\be \label{5.18a}
\hat{G}(\Lambda)f_\gamma=\hat{G}_\gamma(\Lambda) f_\gamma
\ee
for every function $f_\gamma$ cylindrical over a graph $\gamma$ ?\\
2.) If $\hat{G}(\Lambda)$ exists, does its classical limit coincide 
with the classical function $G(\Lambda)$ ?\\
\\
{[}1.]\\
It is easy to see that the first question can be answered affirmatively :\\
Namely, in order that (\ref{5.18a}) holds it is sufficient to show that 
the family of operators $\hat{G}_\gamma(\Lambda)$ is consistently defined.
But this is trivially the case because we defined a function to be 
cylindrical over $\gamma$ if and only if it is a finite linear combination
of spin-network functions which by definition depend non-trivially 
on the holnomy along each of its edges (that is, each edge is labelled 
with a non-trivial irreducible representation of $G$). Thus, if we 
superpose $f_\gamma,f'_{\gamma'}$ with $\gamma\not=\gamma'$ then
$G(\Lambda)[f_\gamma+f'_{\gamma'}]:=   
G_\gamma(\Lambda)f_\gamma+G_{\gamma'}(\Lambda)f'_{\gamma'}$. It is also 
easy to see that this definition leads to the constraint algebra 
\be \label{5.19}   
[\hat{G}(\Lambda),\hat{G}(\Lambda')]
=i\hbar(-\hat{G}([\Lambda),\Lambda'])
\ee
by (\ref{5.18}) since $\hat{G}_\gamma(\Lambda)$ preserves ${\cal H}_\gamma$.
Thus, $\hat{G}(\Lambda)$ exists and defines a consistent quantum 
constraint algebra.\\
{[}2.]\\
To address the second question we first of all notice that we have shown
that 
\be \label{5.20}
G(\Lambda)=\lim_{\gamma\to\Sigma}[\lim_{\hbar\to 0} \hat{G}_\gamma(\Lambda)]
\ee
where the inner bracket has been demonstrated actually only by the
usual ``quantization rule''. A rigorous proof will be given elesewhere 
\cite{46,47}, see also below for a sketch. The outer limit is to be 
understood pointwise on $M$.

What we would like to establish now is the existence of the 
opposite limiting procedure, that is 
\be \label{5.21}
G(\Lambda)=\lim_{\hbar\to 0} 
[\lim_{\gamma\to\Sigma}\hat{G}_\gamma(\Lambda)]
\ee
We will understand the inner bracket to be the operator $\hat{G}(\Lambda)$
defined in (\ref{5.18}) through the self-consistent family of projections 
$\hat{G}_\gamma(\Lambda)$. 

We can then rigorously define the limits 
(\ref{5.20}) and (\ref{5.21}) as follows :\\
Let $\psi^\hbar_{\gamma,m}$ 
be a coherent state, explicitly dependent on Planck's constant, peaked at 
the point $m\in M$ (a smooth field 
configuration) in the following sense : For each graph $\gamma$ and its
associated dual decomposition $P_\gamma$ consider the values of the 
holonomies and momenta $h_e(m),P^e(m)$ respectively. Then the operators
$\hat{h}_e,\hat{P}^e$ have expectation values 
$h_e(m),P^e(m)$ in the state $\psi^m_\gamma$ respectively and satisfy a 
minimal uncertainty condition.

Let now $\gamma_n$ be the family of graphs defined in 
section \ref{s3.2.2}. We then consider the expectation values
\be \label{5.22}
G^\hbar_n(\Lambda,m):=
<\psi^\hbar_{\gamma_n,m},\hat{G}_{\gamma_n}(\Lambda)
\psi^\hbar_{\gamma_n,m}>_{\gamma_n}
\ee
Notice that by definition of the Hilbert space $\cal H$ and the operator
$\hat{G}(\Lambda)$ also
\be \label{5.23}
G^\hbar_n(\Lambda,m)=
<\psi^\hbar_{\gamma_n,m},\hat{G}_{\gamma_n}(\Lambda)\psi^\hbar_{\gamma_n,m}>
\ee
Then the limit (\ref{5.20}) means that
\be \label{5.24}
G(\Lambda,m)=\lim_{n\to\infty}[\lim_{\hbar\to\infty}
G^\hbar_n(\Lambda,m)]
\ee
where now the inner limit is taken at fixed $m,n$ and is meant in the sense
of complex numbers. The limit (\ref{5.21}) on the other hand means that
\be \label{5.25}
G(\Lambda,m)=\lim_{\hbar\to\infty} [\lim_{n\to\infty} G^\hbar_n(\Lambda,m)]
\ee
and will be much more difficult to check for a more general operator because
the $\hbar$ corrections of the inner bracket might not converge.
In our case, however, both limits are immediate and in fact reproduce
$G(\Lambda)$ as we will show as an example in the first publication of
\cite{45}.\\

To conclude, we have shown that there is an anomaly-free quantization of 
the Gauss constraint on the continuum Hilbert space $\cal H$ with the 
corrrect classical limit. The limit (\ref{5.20}) says that the 
regularization procedure is meaningful while the limit (\ref{5.21}) shows 
that the regulator can be removed without picking up divergencies and such 
that we obtain the correct classical limit.

\section{Non-Commutativity Issues}
\label{s6}

The authors of \cite{40} considered the following classical functions
on $(M,\Omega)$ 
\be \label{6.1}
E(S,f):=\int_S  ( \ast E_i f^i)(x)
\ee
where $S$ is an oriented smooth (D-1)-dimensional submanifold of $\Sigma$
and $f^i\in {\cal S}$. Notice that $E(S,f)$
in contrast to our $P(S)$ of (\ref{3.14}) is not gauge covariant
for any choice of $f$ and that $P(S)\not=E(S,f)$ since $P(S)$ depends 
explicitly on both $A$ and $E$ while (\ref{6.1}) depends only on $E$.

In order to compute the Poisson brackets among the $E(S,f)$ induced by
the symplectic structure $\Omega$ one should introduce, as in section
\ref{s3.2.2}, a one parameter family of surfaces $t\mapsto S_t,\;
t\in [-1,1],\;S_0=S$ and smooth regulator functions $g_\epsilon(t),\;
\lim_{\epsilon\to 0} g_\epsilon(t)=\delta(t)$. One obtains 
regulated quantities
\be \label{6.2}
E_\epsilon(S,f):=\int_{-1}^1 dt g_\epsilon(t) \int_{X^{-1}(S_t)} d^2x 
(X^\ast f_a^i E^a_i)(x)
\ee
at the aid of which we compute the Poisson brackets
\be \label{6.3}
\{E(S,f),E(S',f')\}:=\lim_{\epsilon\to 0}
\{E_\epsilon(S,f),E_\epsilon(S',f')\}_\Omega=0
\ee
by (\ref{3.13}). 

The authors of \cite{40} now proceeded as follows :\\ 
Since, according to the symplectic structure of $\Omega$, we formally
have $\{E^a_i(x),A_b^j(y)\}=\delta^a_b \delta^j_i \delta^{(D)}(x,y)$,
they represented the operator $\hat{E}^a_i(x)$ by the functional
derivative $i\delta/\delta A_a^i(x)$ defined on functions of smooth 
connections, substituted this derivative into (\ref{6.1}), applied
it to functions $f_\gamma$ of holonomies of smooth connections over a graph 
$\gamma$ and extended the final operator to distributional   
connections. The result is the following : 

Without loss of generality we can subdivide the graph
sufficiently and orient all the edges of $\gamma$ in such a way 
that any edge of $\gamma$ belongs to one of the following four categories :
i) $e\cap S=\emptyset$, ii) $e\cap S=e$, 
iii) $e\cap S=e(0)$ and $e$ lies on the ``up'' side of $S$ or 
iv) $e\cap S=e(0)$ and $e$ lies on the ``down'' side of $S$.  
Notice that in case iii),iv) the edge is allowed to be tangent at 
$e(0)$. Denote the subset of edges belonging to category iii) and 
iv) respectively by $E^u_S(\gamma)$ and $E^d_S(\gamma)$ respecively
and the subset of vertices in $S\cap (E^u_S(\gamma)\cup
E^d_S(\gamma))$ by $V_S(\gamma)$. Then
\be \label{6.4}
\hat{E}(S,f) f_\gamma=i\hbar\sum_{p\in V_S(\gamma)} \frac{f^i(p)}{2}
[\sum_{e(0)=p,e\in E^u_S(\gamma)} X_i^e
-\sum_{e(0)=p,e\in E^d_S(\gamma)} X_i^e] f_\gamma
\ee
where again $X_i^e$ denotes the right invariant vector field on the 
$e$'th copy of $G$. The expression (\ref{6.4}) defines a self-consistent
family of operators $\hat{E}_\gamma(S,f)$ defined on (a dense subset of)
${\cal H}_\gamma$.

The non-commutativity becomes now obvious 
by choosing for instance $S=S'$ so that
\be \label{6.5}
[\hat{E}(S,f),\hat{E}(S,f')] f_\gamma=\hbar^2\sum_{p\in V_S(\gamma)} 
\frac{f^i(p)f^{j\prime}(p) f_{ij}\;^k}{2}
\sum_{e(0)=p,e\in E^u_S(\gamma)\cup E^d_S(\gamma)} X_k^e] f_\gamma
\ee
and even worse, one cannot write the right hand side
as $\hat{E}(S,[f,f'])$ !

These problems can be overcome as follows :\\
Partition the surface $S$ into disjoint open pieces $S_p$ carrying the 
same orientation as $S$ such that $p$ is 
the only point of $V_S(\gamma)$ lying in $S_p$ and 
$\cup_{p\in V_S(\gamma)} S_p=S$ modulo boundary points. 
For each 
$e\in E^u_S(\gamma)$ or $e\in E^d_S(\gamma)$ respectively,
deform $S_p$ in an arbitrarily small neighbourhood
of $p$ into the direction of $e$ to a surface $S_e$ which intersects $e$ 
transversally in an interior point of $e$ but no other edge of $\gamma$
and which carries the same or opposite orientation as $S_p$.
Obviously, these surfaces qualify as part of a dual decomposition of 
$\gamma$. We can now construct from these data the following function on $M$
which can also be considered as a function on $M_\gamma$ 
\be \label{6.6}
E_\gamma(S,f):=\sum_{p\in V_S(\gamma)} f^i(p)
[\sum_{e(0)=p,e\in E^u_S(\gamma)} P_i^e
-\sum_{e(0)=p,e\in E^d_S(\gamma)} P_i^e]
\ee 
which obviously has classically nothing to do with $E(S,f)$. 
Nevertheless, the results of section \ref{s5} tell us that its quantization
{\it exactly} agrees with (\ref{6.4}), moreover, the algebra of 
operators of this kind reflects precisely the symplectic structure
$\Omega_\gamma$ which is {\it derived} from $\Omega$. 

In conclusion, we have demonstrated that the family of operators 
(\ref{6.4}) can be considered as bona fide quantizations of a family
of classical functions which {\it do not} Poisson commute with 
respect to $\Omega$ and therefore
the apparent contradiction between classical Poisson bracket algebra
and quantum commutator algebra pointed out in \cite{40} evaporates.\\
\\
The discussion of this section seems to reveal that not only there is 
ambiguity in quantizing a given classical functions due to the
always existing possibility to add $\hbar$ corrections, but also vice versa 
that there is an ambiguity in taking the classical limit, in the sense
that one and the same operator can be considered as a quantization of 
two different classical functions. However, this is not the case
if we insist that we begin with a classical phase space and operators
have to have a commutator algebra reflecting the classical Poisson bracket 
algebra. From this point of view, the functions 
(\ref{6.1}) {\it must not} be considered as classical limit of the 
operators (\ref{6.4}) ! One can still argue that the $S^e$ are quite 
arbitrary and that the classical limit is therefore not really well 
defined, but as already said before, a well-defined classical limit
can only be expected in the limit of $\gamma\to\Sigma$ in a definite 
way in which the arbitrariness of the $S^e$ is lost.

Besides, the functions (\ref{6.4}) are unphysical already
from the point of view of the Gauss constraint : it is impossible
to build from them gauge invariant observables except in the limit
of infinitesimal faces where they have been used to build geometrical
operators \cite{12,13,14,15,16}. However, in that limit we get anyway
$E(S,f)\to P(S^e)_i f^i(e(0))$ so that one can equally well construct these
operators from the $P(S^e)_i$ and so there is finally complete agreement
between all the results previously obtained in the literature and our
approach.\\
\\
\\
{\large Acknowledgements}\\
\\
We thank O. Winkler for a careful reading of the manuscript.

\begin{appendix}

\section{The Symplectic Structure for $G=SU(2),U(1)$ as a Two-Form}
\label{sa}

Let us first fix our conventions :\\
For a $p-$form 
$\omega=\omega_{a_1..a_p} dx^{a_1}\otimes..\otimes dx^{a_p}=
\omega_{a_1..a_p} dx^{a_1}\wedge..\wedge dx^{a_p}$
on a finite dimensional manifold $M$ with 
$\omega_{a_1..a_p}=\omega_{[a_1..a_p]}$ we define exterior differential,
interior products with vector fields $v$ and Lie derivatives 
respectively by
\ba 
\label{a.1}
d\omega&=&\partial_a \omega_{a_1..a_p}dx^a\wedge dx^{a_1}\wedge..\wedge
dx^{a_p} \\
\label{a.2}
i_v\;\omega&=&p v^a\omega_{a a_1..a_{p-1}}dx^{a_1}\wedge..\wedge dx^{a_{p-1}}
\\
\label{a.3}
{\cal L}_v\omega &=& [i_v d+d \;i_v]\omega
\ea
Let now $(M,\Omega)$ be a finite dimensional symplectic manifold and 
$f\in C^\infty(M)$. We define the Hamiltonian vector field $\chi_f$ of $f$
by
\be \label{a.4}
i_{\chi_f}\Omega+df=0
\ee 
and the Poisson bracket of $f,g\in C^\infty(M)$ with respect to $\Omega$ by
\be \label{a.5}
\{f,g\}:=-i_{\chi_f} i_{\chi_g}\Omega=\chi_f(g)=i_{\chi_f} dg
\ee 
If $\Omega=d\Theta$ is exact then $\Theta$ is called a symplectic potential
for $\Omega$. Here is a quick method of how to go backwards from 
$\{.,.\}$ to $\Omega$ :\\
Introduce local coordinates $z^\alpha$ on $M$ and corresponding
tensor components $\Omega=\frac{1}{2}\Omega_{\alpha\beta} 
dz^\alpha\wedge dz^\beta$ and define by
$\Omega^{\alpha\gamma}
\Omega_{\gamma\beta}=\delta^\alpha_\beta$ the inverse tensor.
Then the Hamiltonian vecor field of any function $f$ is given by 
$\Omega^{\alpha\beta}\partial_\beta f$. Thus $\{z^\alpha,z^\beta\}=
\Omega^{\gamma\delta}(\partial_\delta z^\alpha)(\partial_\gamma z^\beta)
=-\Omega^{\alpha\beta}$ and so we just have to invert the matrix
of Poisson brackets to obtain $\Omega$.

The reader may verify that with our conventions
the symplectic potential $\Theta=p dq$ of $M=T^\ast\Rl$ leads to
$\{p,q\}=1$.

\subsection{$U(1)$}
\label{sa.1}

The Lie algebra of $U(1)$ is spanned by $i$ (imaginary unit) and is 
therefore Abelian. Let $h\in U(1)$ be a complex number of modulus one. We 
want to compute the symplectic structure $\Omega$ on $M=T^\ast 
U(1)$ corresponding to the brackets $\{h,h\}=0=\{p,p\},\;\{p,h\}=
\frac{i}{2}h$. Let $z^1=p,z^2=h$, then 
$\Omega^{\alpha\beta}=-ih\epsilon^{\alpha\beta}/2$ where
$\epsilon^{\alpha\beta}$ is the completely skew tensor density of 
weight one. Thus $\Omega_{\alpha\beta}=2\epsilon_{\alpha\beta}/(ih)$
and $\Omega=2dp\wedge dh/(ih)=-2i dp\wedge d\ln(h)$ which equals $2 dp\wedge 
d\varphi$ locally if we write $g=\exp(i\varphi)$. $\Omega$ is real and exact 
with symplectic potential $\Theta=-2i p d\ln(h)$.

\subsection{$SU(2)$}
\label{sa.2}

This time there is considerably more work involved and we will only 
sketch the main steps.

Recall the following normalization conditions for our generators
$\mbox{tr}(\tau_i\tau_j)=-2\delta_{ij},\;[\tau_i,\tau_j]=2\epsilon_{ijk}
\tau_k$ (for instance $\tau_j=-i\sigma_j$ the later being the standard 
Pauli matrices). Let us introduce the following global group coordinates
\be \label{a.6}
S:=\mbox{tr}(h),\;T^i:=\mbox{tr}(\tau_i h)
\ee
then $h=(S-T^i\tau_i)/2$ and we have the following relation
$4-S^2=(T^i)^2$. Thus, instead of working with $S,T^i$ we
can work with $\epsilon,T^i$ where $\epsilon=S/|S|=0,\pm 1$ is a discrete 
parameter. From (\ref{3.21a}) -- (\ref{3.21c}) we compute the fundamental 
Poisson brackets 
\ba \label{a.7}
(\Omega^{-1})^{jk}&:=&\{T^j,T^k\}=\{\epsilon,\epsilon\}=\{T^i,\epsilon\}=0 
\nonumber\\
(\Omega^{-1})_j\;^k&:=&\{p_j,T^k\}=-\frac{1}{2}[\epsilon\delta_{jk}
\sqrt{4-(T^m)^2}-\epsilon_{jkl}T^l]\nonumber\\
\{p_j,\epsilon\}&=& 0\nonumber\\
(\Omega^{-1})_{jk}&:=&\{p_j,p_k\}=-\epsilon_{jkl}p_l
\ea
and certainly $(\Omega^{-1})^k\;_j=-(\Omega^{-1})_j\;^k$. Thus, our task 
is to invert the $6$ x $6$ matrix $\Omega^{-1}$ defined in (\ref{a.7}).
We do not worry about the discrete parameter $\epsilon$ which
Poisson commutes with everything in what follows. 

Let us introduce the $3$ x $3$ matrix $\Lambda(v)$ defined
for every vector $v$ by $\Lambda(v)_{ij}:=\epsilon_{ijk} v^k$. Let
also $z^\alpha=T^\alpha,\alpha=1,2,3;\;z^\alpha=p_{\alpha-3},\alpha=4,5,6$
and $\Omega^{\alpha\beta}:=(\Omega^{-1})^{\alpha\beta}
=\{z^\alpha,z^\beta\}$. Then $\Omega^{-1}$ is explicitly given by
\be \label{a.8}
\Omega^{-1}=\frac{1}{2}\left( \begin{array}{cc}
0 & -S 1_3 +\Lambda(T) \\
S 1_3 - \Lambda(T) & 2\Lambda(p)
\end{array} \right)
\ee
Thus, the $6$ x $6$ matrix decomposes into four blocks of $3$ x $3$ matrices.
For the matrix $\Omega$ we now make a similar block matrix ansatz
\be \label{a.9}
\Omega=2\left( \begin{array}{cc}
\Lambda(a) & B \\
-B^T & 2\Lambda(c)
\end{array} \right)
\ee
and study the relations that we obtain from $\Omega\Omega^{-1}=1_6$
for the vectors $a,c$ and the $3$ x $3$ matrix $B$. Using the relations
$\Lambda(U)\Lambda(v)=v\otimes u-(u,v)1_3$ one finds after very lengthy
calculations the result
\be \label{a.10}
\Omega=\frac{1}{2}\left( \begin{array}{cc}
2[\Lambda(p)+\frac{T\otimes p-p\otimes T}{S}] 
& S 1_3-\Lambda(T)+\frac{T\otimes T}{S} \\
-S 1_3-\Lambda(T)-\frac{T\otimes T}{S} & 0
\end{array} \right)
\ee
The matrix (\ref{a.10}) is singular at $S=0$ but we will see that this
is merely a coordinate singularity by simply working out
$\Omega=\frac{1}{2}\Omega_{\alpha\beta}dz^\alpha\wedge dz^\beta$.
The result is 
\be \label{a.11}
\Omega=\frac{1}{2}[(\Lambda(p)+\frac{T\otimes p-p\otimes 
T}{S})_{ij} dT^i\wedge dT^j
+(S 1_3-\Lambda(T)+\frac{T\otimes T}{S})_i\;^j dT^i\wedge dp_j 
\ee
and we find the following global symplectic potential (so $\Omega=d\Theta$ is
exact)
\be \label{a.12} 
\Theta=\frac{1}{2}p_j(\epsilon_{ijk} T_k dT_i-S dT^j+T^j dS)
\ee
from which regularity is obvious. We also find the following locally defined
momentum conjugate to $T^i$ 
\be \label{a.13}
\pi_i=\frac{1}{2}(\epsilon_{ijk} T_k-S\delta_{ij}+\frac{T^i T^j}{S})p_j
\ee
and indeed after lengthy calculations using (\ref{a.7}) we find that
$\{T^i,T^j\}=\{\pi_i,\pi_j\}=0,\;\{p_i,T^j\}=\delta_i^j$. Clearly,
$\pi_i, T^i$ can only be local Darboux coordinates otherwise we would have
displayed $T^\ast SU(2)\equiv T^\ast S^3$ as $T^\ast B^2$ where $B^2$
is a solid ball in $\Rl^3$ (we are missing the discrete information coming 
from $\epsilon$).

The form (\ref{a.12}) could be the starting point of symplectic reduction 
of $(M_\gamma,\Omega_\gamma)$ by the Gauss constraint $G_\gamma(\Lambda)$ 
at the classical level already using methods from geometric quantization
which has not been done so far in the literature to the best of our 
knowledge. However, the manifold $M_\gamma$ reduced by $G_\gamma$ is 
rather singular except in the case of only one copy of $G$ and therefore
is unattractive.

\end{appendix}


\begin{thebibliography}{999}

\parskip -5pt

\bibitem{1} A. Ashtekar, Phys. Rev. Lett. {\bf 57} (1986) 2244,
            Phys. Rev. {\bf D36} (1987) 1587

\bibitem{1a} A. Ashtekar, in ``Mathematics and General Relativity", 
American Mathematical Society, Providence, Rhode Island, 1987

\bibitem{1b} F. Barbero, Phys. Rev. {\bf D51} (1995) 5507\\
F. Barbero, Phys. Rev. {\bf D51} (1995) 5498

\bibitem{1c} T. Thiemann, 
Physics Letters B {\bf 380} (1996) 257-264, gr-qc/960688

\bibitem{2} R. Gambini, A. Trias, Phys. Rev. {\bf D22} (1980) 1380\\
C. Di Bartolo, F. Nori, R. Gambini, A. Trias, Lett. Nuov. Cim.
{\bf 38} (1983) 497\\
R. Gambini, A. Trias, Nucl. Phys. {\bf B278} (1986) 436

\bibitem{3} T. Jacobson, L. Smolin, Nucl. Phys. {\bf B299} (1988) 295

\bibitem{4} A. Ashtekar and C.J. Isham,
Class. Quantum Grav. {\bf 9} (1992) 1433

\bibitem{5} A. Ashtekar, J. Lewandowski, ``Representation
theory of analytic holonomy $C^\star$ algebras'', in ``Knots and
Quantum Gravity'', J. Baez (ed), Oxford University Press, Oxford, 1994\\
A. Ashtekar, J. Lewandowski,
Journ. Geo. Physics {\bf 17} (1995) 191\\
A. Ashtekar and J. Lewandowski, J. Math. Phys. {\bf 36} (1995) 2170

\bibitem{6} D. Marolf and J. M. Mour\~ao, 
``On the support of the Ashtekar-Lewandowski measure'',  
Commun. Math. Phys. {\bf 170} (1995) 583-606

\bibitem{7}
A. Ashtekar, J. Lewandowski, D. Marolf, J.Mour\~ao, T. Thiemann, 
Journ. Math. Phys. {\bf 36} (1995) 6456-6493, gr-qc/9504018

\bibitem{8} C. Rovelli, L. Smolin, Nucl. Phys. {\bf B331} (1990) 80  

\bibitem{8a} R. Gambini, J. Pullin, ``Loops, Knots, Gauge Theories and 
Quantum Gravity'', Cambridge University Press, Cambridge, 1996

\bibitem{9} 
T. Thiemann, 
Journ. Math. Phys. {\bf 39} (1998) 1236-48, hep-th/9601105

\bibitem{10} 
R. De Pietri, C. Rovelli, 
Phys. Rev. {\bf D54} (1996) 2664\\
R. De Pietri, Class. Quantum Grav. {\bf 14} (1997) 53

\bibitem{11} 
T. Thiemann,
Journ. Math. Phys. {\bf 39} (1998) 3372-92, gr-qc/9606092

\bibitem{12} C.\ Rovelli, L.\ Smolin, 
Nucl. Phys. B {\bf 442} (1995) 593; Erratum :
Nucl. Phys. B {\bf 456} (1995) 734

\bibitem{13} A. Ashtekar, J. Lewandowski, 
Class. Quantum Grav. {\bf 14} A55-81 (1997)

\bibitem{14}
R. Loll, Nucl. Phys. {\bf B460} (1996) 143\\
R. Loll, Class. Quantum Grav. {\bf 14} (1997) 1725\\
R. Loll, Nucl. Phys. {\bf B500} (1997) 405

\bibitem{15} A. Ashtekar, J. Lewandowski, 
Adv. Theo. Math. Phys. {\bf 1} (1997) 388-429 

\bibitem{16} 
T. Thiemann, 
Journ. Math. Phys. {\bf 39} (1998) 3347-71, gr-qc/9606091

\bibitem{17} 
T. Thiemann, 
Class. Quantum Grav. {\bf 15} (1998) 839-73, gr-qc/9606089
  
\bibitem{18} 
T. Thiemann, 
Class. Quantum Grav. {\bf 15} (1998) 875-905, gr-qc/9606090

\bibitem{19} 
T. Thiemann, 
Class. Quantum Grav. {\bf 15} (1998) 1249-80, gr-qc/9705018

\bibitem{20} 
T. Thiemann, 
Class. Quantum Grav. {\bf 15} (1998) 1281-1314, gr-qc/9705019
 
\bibitem{21} 
T. Thiemann, 
Class. Quantum Grav. {\bf 15} (1998) 1487-1512, gr-qc/9705021

\bibitem{22}
T. Thiemann,
Class. Quantum Grav. {\bf 15} (1998) 1463-85, gr-qc/9705020

\bibitem{23} 
T. Thiemann,
Class. Quantum Gravity {\bf 15} (1998) 1207-47, gr-qc/9705017

\bibitem{23f} T. Thiemann, ``Quantum Spin Dynamics (QSD) : VIII. The 
Classical Limit'', to appear

\bibitem{23a} M. Reisenberger, C. Rovelli,
Phys. Rev. {\bf D56} (1997) 3490-3508

\bibitem{23b} J. Baez,
Class. Quantum Grav. {\bf 15} (1998) 1827-58

\bibitem{23c} F. Markopoulou, L. Smolin,
Phys. Rev. {\bf D58} (1998) 084032

\bibitem{23d} J. W. Barret, L. Crane, Class. Quantum Grav. 14 (1997) 2113

\bibitem{23e}
L. Freidel, K. Krasnov,
Adv. Theor. Math. Phys. {\bf 2} (1999) 1183

\bibitem{24} 
K. Krasnov, Gen. Rel. Grav. 30 (1998) 53-68

\bibitem{25} 
A. Ashtekar, C. Beetle, S. Fairhurst, 
Class. Quantum Grav. {\bf 17} (2000) 1317


\bibitem{26} A. Ashtekar, A. Corichi, K. Krasnov, 
``Isolated horizons : The classical phase space'', 
gr-qc/9905089

\bibitem{27} 
A. Ashtekar, C. Beetle, S. Fairhurst, Class. Quantum 
Grav. {\bf 16} (1999) L1-L7, gr-qc/9812065

\bibitem{28} A. Ashtekar, K. Krasnov, ``Quantum Geometry and Black Holes'', 
gr-qc/9804039

\bibitem{29} 
A. Ashtekar, J. Baez, A. Corichi, K. Krasnov, Phys. Rev. Lett. 
{\bf 80} (1998) 904

\bibitem{30} A. Ashtekar, ``Interface of General Relativity, Quantum 
Physics amd Statistical Mechanics : Some Recent Developments'',
gr-qc/9910101

\bibitem{31} S.A. Fulling, ``Aspects of Quantum Field Theory in Curved
Space-Time'', Cambridge University Press, Cambridge, 1989


\bibitem{46} T. Thiemann, ``Gauge Field Theory Coherent States (GCS) 
: I. General Properties'', hep-th/0005233

\bibitem{47} 
T. Thiemann, O. Winkler, ``Gauge Field Theory Coherent States (GCS) : II.
Peakedness Properties'', hep-th/0005237 \\
T. Thiemann, O. Winkler, ``Gauge Field Theory Coherent States (GCS) : III.
Ehrenfest Theorems'', hep-th/0005234\\
T. Thiemann, ``Gauge Field Theory Coherent States (GCS) : V.
Extension to Higgs Fields and Fermions'', to appear

\bibitem{45} 
T. Thiemann, O. Winkler, ``Gauge Field Theory Coherent States (GCS) : IV.
Infinite Tensor Product and Thermodynamic Limit'', hep-th/0005235\\ 
H. Sahlmann, T. Thiemann, O. Winkler, ``Gauge Field Theory Coherent States 
(GCS) : VI. Photons and Gravitons Propagating on Quantum Spacetimes'',
to appear\\ 
H. Sahlmann, T. Thiemann, O. Winkler, ``Gauge Field Theory Coherent States 
(GCS) : VII. The Non-Perturbative $\gamma$-Ray Burst Effect'', to appear 

\bibitem{40} A. Ashtekar, A. Corichi, J.A. Zapata, 
Class. Quantum Grav. {\bf 15} (1998) 2955

\bibitem{32} J. Baez, S. Sawin, ``Functional Integration on Spaces of 
Connections", q-alg/9507023\\
J. Baez, S. Sawin, ``Diffeomorphism Invariant Spin-Network States", 
q-alg/9708005

\bibitem{33}
J. Lewandowski, T. Thiemann, 
Class. Quantum Grav. {\bf 16} (1999) 2299-2322, gr-qc/9901015

\bibitem{34} C.\ Rovelli, L.\ Smolin,
``Spin networks and quantum gravity'' pre-print CGPG-95/4-1,
Phys. Rev. {\bf D52} (1995) 5743

\bibitem{35} J.\ Baez, 
Adv. Math. {\bf 117} (1996) 253-272

\bibitem{36} A. Rendall, 
Class. Quantum Grav. {\bf 10} (1993) 605 \\       
A. Rendall, ``Adjointness Relations as a Criterion for Choosing an Inner
Product'', gr-qc/9403001

\bibitem{37} E. Hebey, ``Sobolev Spaces on Riemannian manifolds'',
Lecture Notes in Mathematics {\bf 1635}, Springer Verlag, Berlin, 1996

\bibitem{38} Y. Choquet-Bruhat, C. DeWitt-Morette, ``Analysis, Manifolds 
and Physics", part I, North Holland, Amsterdam, 1989

\bibitem{39} S. Lang, ``Differential Manifolds'', Addison Wesley,
Reading, Massachusetts, 1972

\bibitem{39a} H. Seifert, W. Threlfall, ``Lehrbuch der Topologie'', Chelsea
Publishing Company, New York, 1980

\bibitem{42} 
J. Kogut, L. Susskind, Phys. Rev. D {\bf 11} (1975) 395\\
P. Renteln, L. Smolin, Class. Quantum Grav. {\bf 6} (1989) 275\\
R. Loll, Nucl. Phys. Proc. Suppl. {\bf 57} (1997) 255\\
R. Loll, Class. Quantum Grav. {\bf 15} (1998) 799\\
R. Loll, gr-qc/9805049, http://www.livingreviews.org/.

\bibitem{43} N. J. Vilenkin, ``Special functions and the theory of group 
representations", American Mathematical Society, Povidence, Rhode
Island, 1968

\bibitem{44} Y. Yamasaki, ``Measures on Infinite Dimensional Spaces'',
World Scientific, Singapore, 1985


\end{thebibliography}
\end{document}